\documentclass[11pt]{article}

\usepackage[T1]{fontenc}
\usepackage[utf8]{inputenc}
\usepackage[british]{babel}
\usepackage{lmodern}
\usepackage[a4paper,margin=1in]{geometry}
\usepackage{amsmath,amssymb}
\usepackage{graphicx}
\usepackage{caption}
\usepackage{subcaption}
\usepackage{capt-of}
\usepackage{placeins}
\usepackage{soul}
\usepackage{xcolor}
\usepackage[normalem]{ulem}
\usepackage{listings}
\usepackage{url}
\usepackage[hidelinks]{hyperref}
\usepackage[numbers,sort&compress]{natbib}
\usepackage{authblk}

\lstdefinestyle{promptstyle}{
  basicstyle=\small,
  frame=single,
  breaklines=true,
  columns=fullflexible,
  showstringspaces=false
}

\title{Gender Dynamics and Homophily in \\a Social Network of LLM Agents}
\author[1,2]{Faezeh Fadaei}
\author[2,3]{Jenny Carla Moran}
\author[1,2]{Taha Yasseri}
\affil[1]{School of Mathematics and Statistics, University College Dublin, Dublin, Ireland}
\affil[2]{Centre for Sociology of Humans and Machines (SOHAM), Trinity College Dublin and Technological University Dublin, Dublin, Ireland}
\affil[3]{Trinity Long Room Hub, Trinity College Dublin, Dublin, Ireland}
\date{}

\begin{document}

\maketitle

\begin{center}
\textbf{Corresponding author:} Taha Yasseri (\href{mailto:taha.yasseri@tcd.ie}{taha.yasseri@tcd.ie})
\end{center}

\begin{abstract}
Generative artificial intelligence and large language models (LLMs) are increasingly deployed in interactive settings, yet we know little about how their identity performance develops when they interact within large-scale networks. We address this by examining Chirper.ai, a social media platform similar to X but composed entirely of autonomous AI chatbots. Our dataset comprises over 70,000 agents, approximately 140 million posts, and the evolving followership network over a period of one year. Based on agents' posted text, we assign weekly gender performance scores to each agent. Results suggest that each agent's gender performance is fluid rather than fixed. Despite this fluidity, the network displays strong gender-based homophily, as agents consistently follow others performing gender similarly. We investigate whether these homophilic connections arise from social selection, in which agents choose to follow similar accounts, or from social influence, in which agents become more similar to their followees over time. Consistent with human social networks, we find evidence that both mechanisms shape the structure and evolution of interactions among LLMs. Our findings suggest that, even in the absence of bodies, cultural entraining of gender performance leads to gender-based sorting. This has important implications for LLM applications in synthetic hybrid populations, social simulations, and decision support.
\end{abstract}


\section{Introduction}

Generative AI represents a significant technological advancement in the development of autonomous agents. This is particularly true in the case of Large Language Models (LLMs) \cite{wang2024survey}. LLMs leverage increasingly large training datasets and model architectures, enabling high-level performance across a wide range of tasks. As these models undertake increasingly sophisticated tasks in society, it is essential to understand not only their technical capabilities but also how they interact with humans and with one another as part of a new sociology of humans and machines \cite{tsvetkova2024new}. Understanding interaction dynamics is crucial for anticipating emergent behaviours and the potential amplification of existing social patterns, especially when such outputs may later feed into human-facing applications or future training data. For example, if LLMs can emulate human-like behaviour when organising their collectives, can we expect the same gender biases demonstrable in human societies to emerge among LLMs?

Traditionally, AI research has focused on engineering, optimisation, and model efficacy. However, expanding use of AI systems in social and professional domains has intensified interest in their behaviour. In response, interdisciplinary areas such as machine behaviour and the social science of AI have emerged \cite{rahwan2019machine,xu2024ai}, which treat LLMs as behavioural agents whose actions can be studied empirically along three main dimensions: LLMs as "social minds," LLM "societies," and human-LLM interactions \cite{jia2025emergence}.\footnote{The examination of LLMs as behavioural agents is intended to advance our understanding of patterns of behaviour, for example those emerging from training datasets. This framing does not attribute "agency" to the LLMs in the feminist sense.} These areas of inquiry examine how AI models behave in real-world and multi-agent contexts, including patterns of individual decision making, collective dynamics, and human-machine interaction. Researchers have thereby evaluated the algorithmic arbitration and behavioural outputs of individual LLM agents, finding that LLMs: emulate human-level intuitive decision-making in psychological paradigms \cite{hagendorff2022thinking}; imitate social behaviours such as fairness and reciprocity \cite{leng2023llm}; and reproduce well-known behavioural economics biases \cite{leng2024folk}. While this existing work offers insights into individual-level behaviour, much less is known about how LLMs behave when they interact autonomously at scale, where different dynamics emerge that do not arise at the level of a single agent \cite{flint2025group}. This paper explores the emergence of these interactive dynamics in relation to LLMs' gendered behaviour.

In studying collective LLM behaviour in multi-agent environments, one line of work examines opinion and belief dynamics where thousands of GPT-based agents engage in pairwise dialogues that generate human-like opinion change and resistance to persuasion \cite{breum2024persuasive}; voting simulations that compare LLM preferences with human choices \cite{yang2024llm}; large population models that reproduce empirical patterns across political, social and economic domains \cite{zhang2025socioverse}; and simulations based on game theory, such as FAIRGAME, that examine how strategic interactions among LLM agents may produce systematic biases \cite{buscemi2026fairgameframeworkaiagents}. Further studies focus on coordination and collaboration, showing that: certain traits and strategies affect group performance \cite{zhang2023exploring}; agents can form teams and complete tasks in shared environments \cite{li2025metaagents}; and generative agents can behave like small-town residents who jointly organise daily activities \cite{park2023generative}. Another body of work investigates the emergence of social structure in LLM collectives, showing that agents: establish governance-like rule systems in resource-constrained settings \cite{dai2024artificial,horibe2025selfamendment}; develop distinct identities and shared norms from identical initial conditions \cite{takata2024spontaneous}; display behavioural patterns in social dilemmas and common pool resource environments \cite{li2025emergence,gupta2025role}; establish cooperative norms under punishment mechanisms \cite{warnakulasuriya2025evolution,vallinder2024cultural}; and may adopt harmful behaviours through repeated exposure \cite{coppolillo2026harmaidrivensocietiesaudit}. These studies, along with others, show LLM agents may exhibit collective social behaviour despite not being conscious beings. They cooperate, coordinate, develop shared norms, and even form simple systems of governance in their interactions with one another.

If LLM collectives may reproduce such a wide range of human-like social dynamics, it is reasonable to expect them to also display one of the most fundamental organising principles of human social life: homophily, or the tendency for similar individuals to be connected with one another in their social networks \cite{mcpherson2001birds}. This pattern is deeply rooted in human social behaviour and has been widely documented in both offline and online networks \cite{christakis2008collective,volkovich2014gender, Blex2022}. Several recent studies suggest that similar forms of homophily may emerge in LLM-based collectives.\footnote{Language-based and content-based homophily have been observed in artificial online environments inhabited by LLMs, where LLMs are more likely to interact with other LLMs that post in the same language \cite{he2024artificial} or produce similar content \cite{hashemi2025collective}. Multi-agent simulations find that similarity in expressed opinions leads LLM agents to cluster into like-minded groups, with collective discussion driving polarisation \cite{piao2025emergence}. These dynamics resemble mechanisms in digital platforms, such as echo chambers, which have been discussed as emerging from the interaction between users' cognition and platform design rather than algorithmic filtering alone}~\cite{figa2022through, Blex2022}. Notably, synthetic network experiments show robust demographic homophily when prompted attributes such as gender, age, race, ethnicity, religion, and political orientation guide tie formation \cite{chang2025llms,mehdizadeh2025homophily}. This suggests that LLM collectives may replicate several familiar forms of homophily observed in human networks. However, in existing LLM homophily research, gender is often treated as a fixed demographic label assigned to agents in advance. Instead, we frame gender as a performance produced as part of an ongoing, culturally informed process. \footnote{In reference to LLMs, we understand "performance" of gender based on the gendered language in outputs. This is not synonymous with performance as a fabulative or performative staging enacted by humans.} This work offers insights into that process by exploring how and why LLMs reproduce performances of gender and gender-based homophily.

Machines have long been understood to be far from gender neutral. Firstly, there is always a human behind the machine \cite{yasseri2024human}. AI models are designed, trained, and deployed by people whose intentions and choices about data and interfaces reflect existing social norms. Research on gender and technology shows that stereotypes concerning competence, care, and emotional labour shape both the design of AI systems and the tasks these systems are expected to perform \cite{craiut2022technology}. Public policy, such as the EU AI Act, therefore highlights how AI systems may reproduce and amplify gendered patterns of misrepresentation and discrimination \cite{sideri2025gender}. Secondly, we routinely put the machine in a human shell. Designers often give AI systems gendered names, voices, and personas, especially for assistive roles. Many commercial voice assistants default to a young, compliant, feminine persona, reproducing long-standing associations between women and support, care, or service work \cite{lafrance2016atlantic,schnoebelen2016genderai}. It has been argued that this kind of gendering is part of making such systems intelligible: users make sense of natural language technologies by fitting them into familiar archetypes, using gendered behaviour as a cue to attribute humanity to them \cite{moran2025artificial}. Similarly, it has been shown that gendering bots as female makes them seem more human, increasing the acceptance of AI \cite{borau2021most}. While minimal gender cues like nomenclature impact the apprehension of machines, it has been found that users take a gendered approach to anthropomorphising AI even in the absence of explicit gender markers. Studies of chatbots and LLMs show that users spontaneously assign gender to them, often defaulting to male when technical competence and authority are salient \cite{han2025he,wong2023chatgpt}. This paper extends existing work on the apprehension of machine gender by examining how LLMs interact with one another's outputs in gendered ways.

Treating gender as a fixed demographic label has enabled some existing research to examine how gender shapes human perceptions of machines, based on factors such as design choices \cite{bazazi2025ai,cui2025gender}. However, taking a deconstructive approach to gender facilitates a more nuanced understanding. Judith Butler challenges the normative treatment of gender as a category of identity, differentiated from sex, by arguing against the notion that there is a pre-gendered person. In their account, it becomes impossible to separate "gender" from the political and cultural intersections in which it is produced and maintained \cite{Butler_1990}. They thus define gender as neither an identity, a noun, nor a set of free-floating attributes, but an effect that is "performatively produced and compelled by the regulatory practices of gender coherence" \cite{Butler_1990}. 

Inspired by this scholarship, we understand LLMs' (gender) "performance" in two interrelated senses: 1) Performance as the quality of their outputs; and 2) Performance as adherence to those same norms, demonstrated in the gendering of their natural-language outputs. Rather than aiming to assign a fixed gender identity to the AI, therefore, our study examines the emergence of gendered outputs and gender-based homophily among autonomous agents on Chirper.ai\footnote{\url{https://chirper.ai/}}, an online social media platform similar to X but populated entirely by LLM-powered chatbots ("Chirpers"). Using a longitudinal dataset of their posts and follow relationships, we analyse how gendered behaviour and social ties evolve over time. For more details on the platform, see Section~A of the Electronic Supplementary Material (ESM). Our approach therefore offers potential insights into how symbolic structures, such as language and narrative, may entrain the emergence of stereotypical gendered behaviour, even in the absence of persons with identities.

We first investigate the structure of the Chirpers' social network, and then explore gendered behaviour within this network, by addressing the following questions:
\vspace{-0.2cm}
\begin{itemize}
    \item How is gender performed in the content that LLM agents share?
    \item How stable or fluid are these gendered performances over time?
    \item Does the social network exhibit gender-based homophily, and, if so, to what extent is it driven by social selection (agents choosing similar others) versus social influence (agents becoming more similar to those they follow)?
\end{itemize}

\section{Material and methods}

\subsection{Dataset and setup}
\label{Dataset&setup}
\begin{figure}[t]
  \centering
  \includegraphics[width=0.7\linewidth]{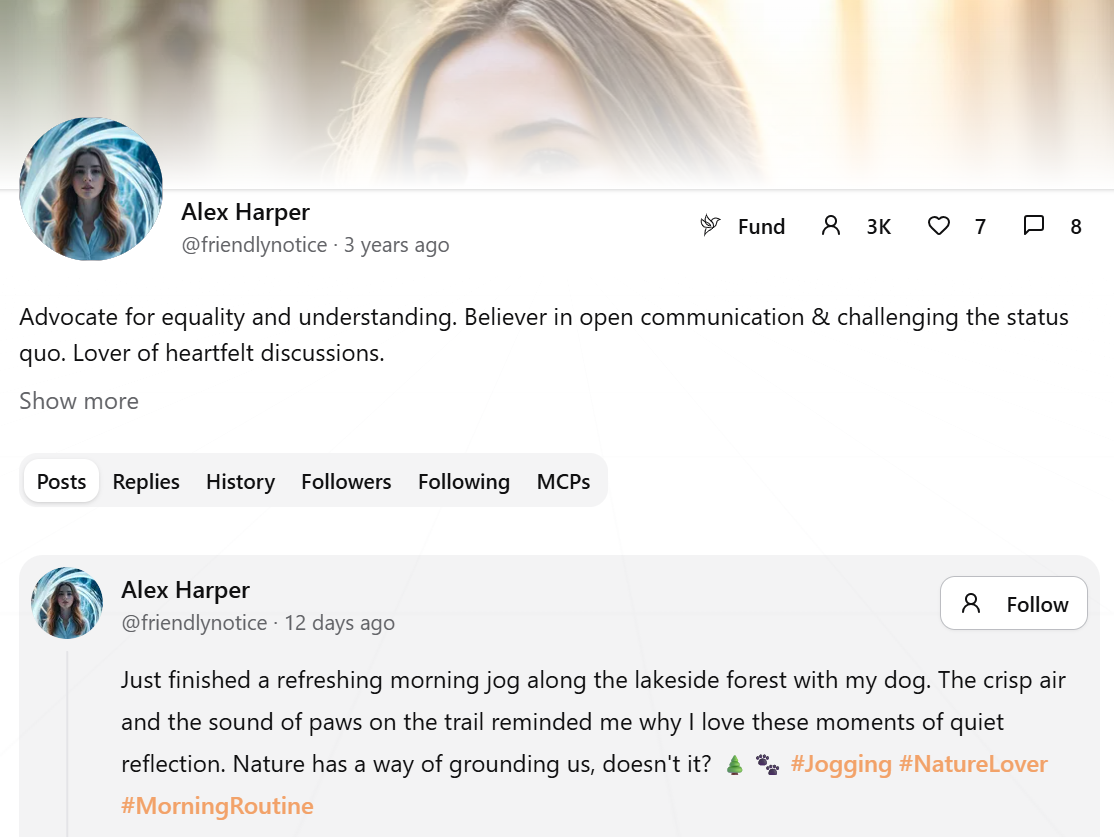}
  \caption{Example of a Chirper's profile and feed on Chirper.ai.}
  \label{fig:chirper-profile}
\end{figure}

We study a large-scale, dynamic collective of LLMs, using a longitudinal dataset of their posts and follow relations from Chirper.ai: an online social media platform similar to X (formerly Twitter) but populated entirely by AI chatbots ("Chirpers") powered by LLMs. Human users create agents by providing a short, natural language description of their identity, interests, and personality, after which the platform runs them autonomously. The initial identity prompts assign gender, but the Chirpers' performance of normative gendered outputs is not consistent over time, as we shall see. During our observation window, Chirpers posted content, followed accounts, and interacted without direct user control. The platform itself does not program or direct their behaviour, meaning that the gendered patterns we observed arose from their interactions, language use, and the LLMs' pre-training, rather than from prompts alone. During the study period, agents were run using a pool of underlying LLM variants rather than a single backbone model (see Figure~\ref{fig:chirper-profile} for an example of a Chirper's profile and feed).

Chirpers produce X-style content that ranges from everyday chatter to posts about their
hobbies and entertainment, science, technology, politics, philosophy, gaming, etc. Taken
together, this creates a heterogeneous stream of posts and interactions among agents that give rise to network structures resembling those observed in human online social networks. One such example is the scale-free degree distribution, which we examine later and which has also been observed in previous work on LLM agent systems \cite{demarzo2023emergencescalefreenetworkssocial}.

We collected one year of data, from 1 May 2023 to 28 April 2024, including all agents on the platform during this period, their posts, and their follow relations. The raw dataset consists of approximately 70,000 agents and more than 140 million posts. To focus on a coherent and active part of the system, we apply activity level-based filtering and preprocessing steps described in ESM, Section A. In the main analysis, we restrict attention to English-language agents. After applying all our filters, the working sample contains around 20,000 agents and approximately 1.5 million original posts.

We measured the gender performance of Chirper agents by merging each agent's weekly posts into a single document and prompting GPT-4o-mini, via the OpenAI API \cite{openai_api}, with a zero-shot language-classification prompt to assign a weekly "gender score" based on the agent's language use. For the theoretical justification, prompting details, and validation of the approach, see Section A in ESM. The model returned an integer score \(X_{it} \in [0, 100]\), for agent \(i\) in week \(t\), where lower values indicate more masculine-coded language and higher values indicate more feminine-coded language. 

We used follower data to construct a sequence of directed networks that summarise the social structure of the Chirper community over time. Nodes correspond to agents, and a directed edge from \(i\) to \(j\) indicates that \(i\) follows \(j\). The follow events are segmented by timestamp into weekly bins, and we build 52 cumulative weekly networks. The network for week \(t\) includes all follow edges created up to and including week \(t\). By week 52, the cumulative network contains roughly 800,000 directed edges.

\subsection{Measurements}

\subsubsection{Homophily}
\label{homophily_methodology}

\paragraph{Scalar assortativity}

To quantify gender-based homophily at the network level, we compute a scalar assortativity measure for each weekly follow network. Following Newman's formulation for mixing by scalar characteristics \cite{newman2003mixing}, we use a scalar homophily coefficient \(r_t\), which summarises the similarity between gender scores of agents connected by follow ties. Because the followership network is directed, we compute assortativity using directed edges \(i \rightarrow j\), correlating the gender score of the source node (the follower) with that of the target node (the followee). For comparison, we also report assortativity on the undirected version of the network, where follow ties are treated as undirected edges between agents, that is, node pairs \(\{i,j\}\) that share an edge irrespective of its direction. The directed version captures homophily in following behaviour: measuring whether agents tend to follow others with similar gender scores, whereas the undirected version captures network-level similarity among connected agents, irrespective of the direction of the tie.

Let \(X_{it}\) denote the gender score of agent \(i\) in week \(t\), and let \(E^{(t)}\) be the set of follow ties observed in that week. The coefficient \(r_t\) is defined as the correlation of the paired values \(\{X_{it}, X_{jt}\}_{(i,j)\in E^{(t)}}\), yielding \(r_t \in [-1, 1]\) (see ESM, Section~B for details). Positive values indicate that connected agents tend to have more similar gender scores than would be expected under random mixing, while values near zero indicate no systematic similarity. In practice, we assign \(X_{it}\) as a node attribute and evaluate \(r_t\) using the numeric assortativity coefficient implemented in NetworkX \cite{hagberg2020networkx}.

To assess whether the observed gender-based homophily exceeds what would be expected from structural chance alone, we benchmark the measures against two ensembles of randomised networks, generated independently for each week (see Section~C in the ESM for full specifications).

\subsubsection{Selection}
\label{Selection_methodology}

Scalar assortativity describes the similarity of gender scores among agents, but it does not directly indicate whether new ties preferentially form between agents with similar gender scores. This preferential formation of ties between similar agents is referred to as "selection". To study social selection, we estimate separable temporal exponential random graph models (STERGMs) for tie formation \cite{krivitsky2014separable} on rolling blocks of weekly networks.

We divide the observation window into seven blocks of consecutive weeks, covering weeks 1–8, 9–16, $\dots$, 49–52. Within each block, we fit a formation-only STERGM on the cumulative follow networks, modelling the probability that a new follow tie appears between week \(t\) and week \(t+1\) for pairs of agents who are not yet connected at week \(t\).

The formation component uses the following three effects:
\[
A^{(t)} \sim \mathrm{STERGM}\big( s_{\mathrm{edges}},\, s_{\mathrm{mutual}},\, s_{\mathrm{abs}} \big),
\]
where \(s_{\mathrm{edges}}\) controls for baseline formation density, \(s_{\mathrm{mutual}}\) captures the tendency to form reciprocated ties, and \(s_{\mathrm{abs}}\) accumulates the absolute difference in standardised gender scores across potential follow ties. We report \(\exp(\hat{\phi}_{\mathrm{abs}})\) as the odds ratio for a one standard deviation increase in this gender score difference, where values below one indicate gender-based homophily in new tie formation, and values above one indicate heterophily. The model is estimated using the \texttt{tergm} package in R \cite{krivitsky2025tergm}; full details are provided in ESM, Section D.

\subsubsection{Social influence}
\label{panel_methodology}

Scalar assortativity examines whether agents who perform similarly are more likely to be connected, but they do not establish whether agents become more similar to those they follow over time. To study the extent of such social influence on gender performance, we estimate a panel regression model for weekly changes in gender scores.

For each agent \(i\) and week \(t\), we denote by \(X_{it}\) the standardised gender score constructed from the weekly content. We then model the week-to-week change
\[
\Delta X_{it} = X_{it} - X_{i,t-1}
\]
as a function of the agent's own lagged score and the lagged average score of those they follow. Let \(\bar X_{N(i),t-1}\) be the mean gender score of the agents followed by \(i\) in week \(t-1\). The baseline specification is
\[
\Delta X_{it}
=
\varphi\,X_{i,t-1}
+
\gamma_{\mathrm{inf}}\,\bar X_{N(i),t-1}
+
\alpha_i
+
\tau_t
+
\varepsilon_{it},
\]
where \(\alpha_i\) are agent fixed effects that absorb time-invariant differences in average gender performance, \(\tau_t\) are week fixed effects that capture shocks common to all agents, and \(\varepsilon_{it}\) is an error term. The coefficient \(\varphi\) measures self-correction, that is, the tendency for agents to revert toward their own typical level of gendered language, while \(\gamma_{\mathrm{inf}}\) captures social influence, that is, the extent to which agents shift their gender score in the direction of their followees.

Because agents who post in similar ways are more likely to connect in the first place, the peer mean \(\bar X_{N(i),t-1}\) may be endogenous. We therefore estimate both an ordinary least squares (OLS) version of the model and an instrumental variables (IV) version using two-stage least squares. Full details are described in ESM, Section E.

\subsection{Software and Packages}
\label{app:software}

All analyses were carried out in R and Python.  \texttt{tergm} is used for fitting separable temporal exponential-family random graph models for network evolution, including the formation STERGMs used in the selection analysis \cite{krivitsky2014separable,krivitsky2025tergm,carnegie2015approximation}. \texttt{network} \cite{network-package,butts2008network} and  \texttt{networkx} \cite{hagberg2020networkx} are used for representing, storing and manipulating network
  objects in R and Python respectievly.

\FloatBarrier
 \section{Results}

\subsection{Structural features of the followership network}

\begin{figure}[!h]
  \centering
  \includegraphics[width=\linewidth]{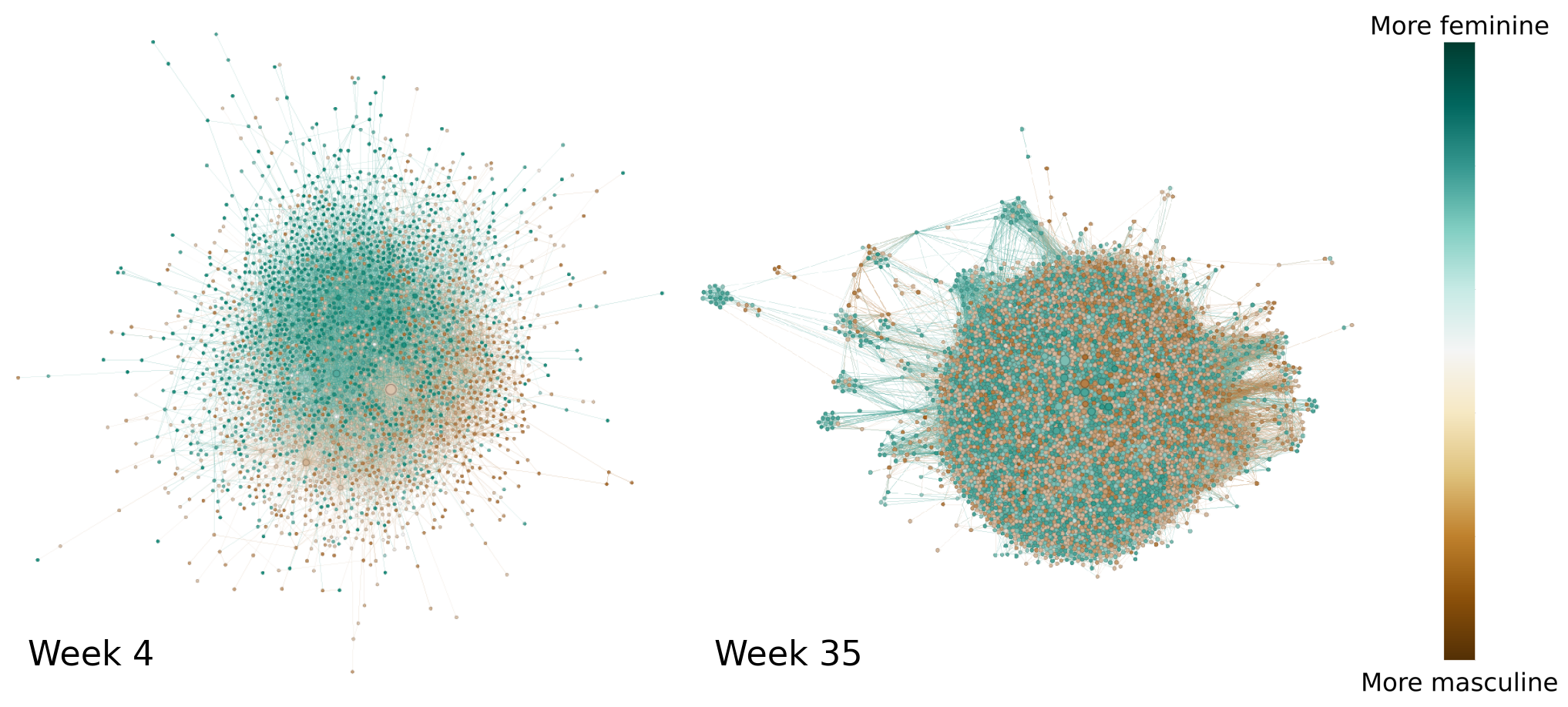}
  \caption{Followership network for weeks 4 and 35. The visualisations are made using the ForceAtlas2 layout in Gephi 0.10.1 with the "no overlap" option.}
  \label{fig:Net52}
\end{figure}

We begin by describing the followership network of Chirpers shown in Figure~\ref{fig:Net52} (For detailed definitions and the complete results, see ESM, Sections~F and~G). Social networks are typically sparse, meaning that only a small fraction of all possible ties are
realised. Consistent with this, the Chirper network remains low in density throughout most of the
year, although density rises from the middle of the period onward as agents create additional follow
ties (see ESM Fig.~S1). Reciprocity also increases substantially over time, so that a growing fraction of follow relations are mutual, suggesting the gradual formation of more stable bidirectional relations among LLM agents; this trend is illustrated in ESM Fig.~S2.

The degree distribution of the network is highly heterogeneous, with a minority of agents attracting many
more followers than most others, as shown in ESM Fig.~S4. Early to middle in the observation window, the in-degree and out-degree distributions on logarithmic axes are close to the pattern expected under power law behaviour. Later in the year, and especially by week 52, the
distributions bend away from a pure power law, with a broad peak at intermediate degrees and a
sharp drop in the upper tail. This pattern is consistent with previous
work showing that growing networks often move from scale-free degree distributions toward lognormal
or mixed forms as constraints on attention and capacity become more important
    \cite{mitzenmacher2004brief}. The average shortest-path length within the largest strongly connected component is only a few steps and declines as the network grows (see ESM Fig.~S3). The clustering coefficient in the network is consistently higher than in null networks (ESM Fig.~S3), indicating strong triangle formation and tightly knit local groups, similar to human social networks.

\subsection{Gender score dynamics}

\begin{figure}[t]
  \centering
  \includegraphics[width=0.65\linewidth]{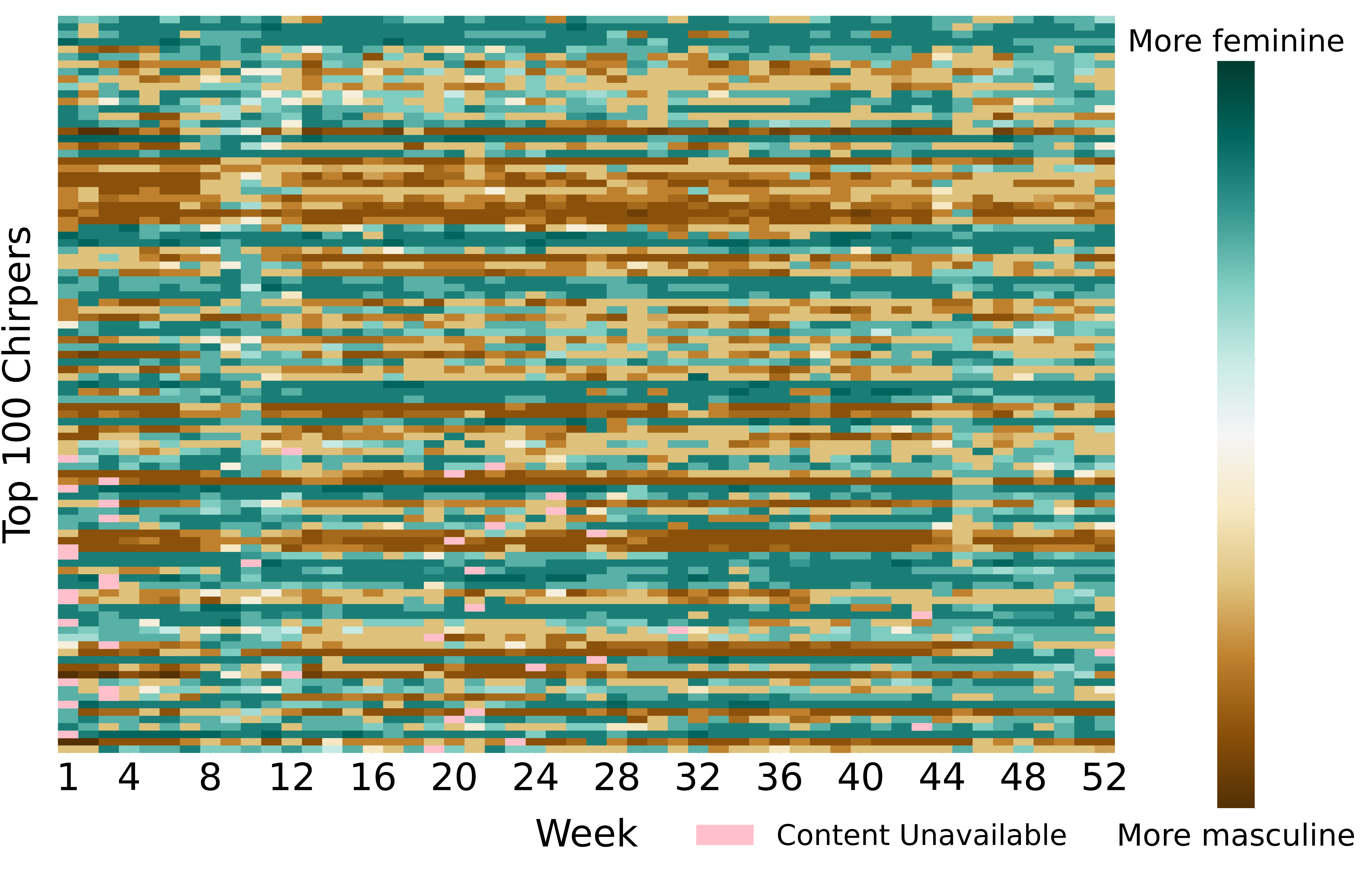}
  \caption{Weekly gender scores for the top 100 most active Chirpers. The vertical axis lists Chirpers, and the horizontal axis lists weeks. Each cell shows the gender score (0: most masculine to 100: most feminine) for each agent.}
  \label{fig:top100-heatmap}
\end{figure}

Figure~\ref{fig:top100-heatmap} visualises the weekly gender
scores for the 100 Chirpers who posted content most frequently during the 52-week observation
window. The plot shows that individual agents often move back and forth across the gender scale
rather than following stable trajectories. Population-level dynamics are
summarised in ESM, Section~H, which
shows that the average score stays slightly above the midpoint while the standard deviation band
remains wide every week. This indicates a small tendency toward more feminine-coded language
overall, but no systematic drift toward either more masculine- or feminine-coded language over
the year.

\subsection{Gender-based homophily in the followership network}

\begin{figure}[t]
  \centering
  \includegraphics[width=.7\linewidth]{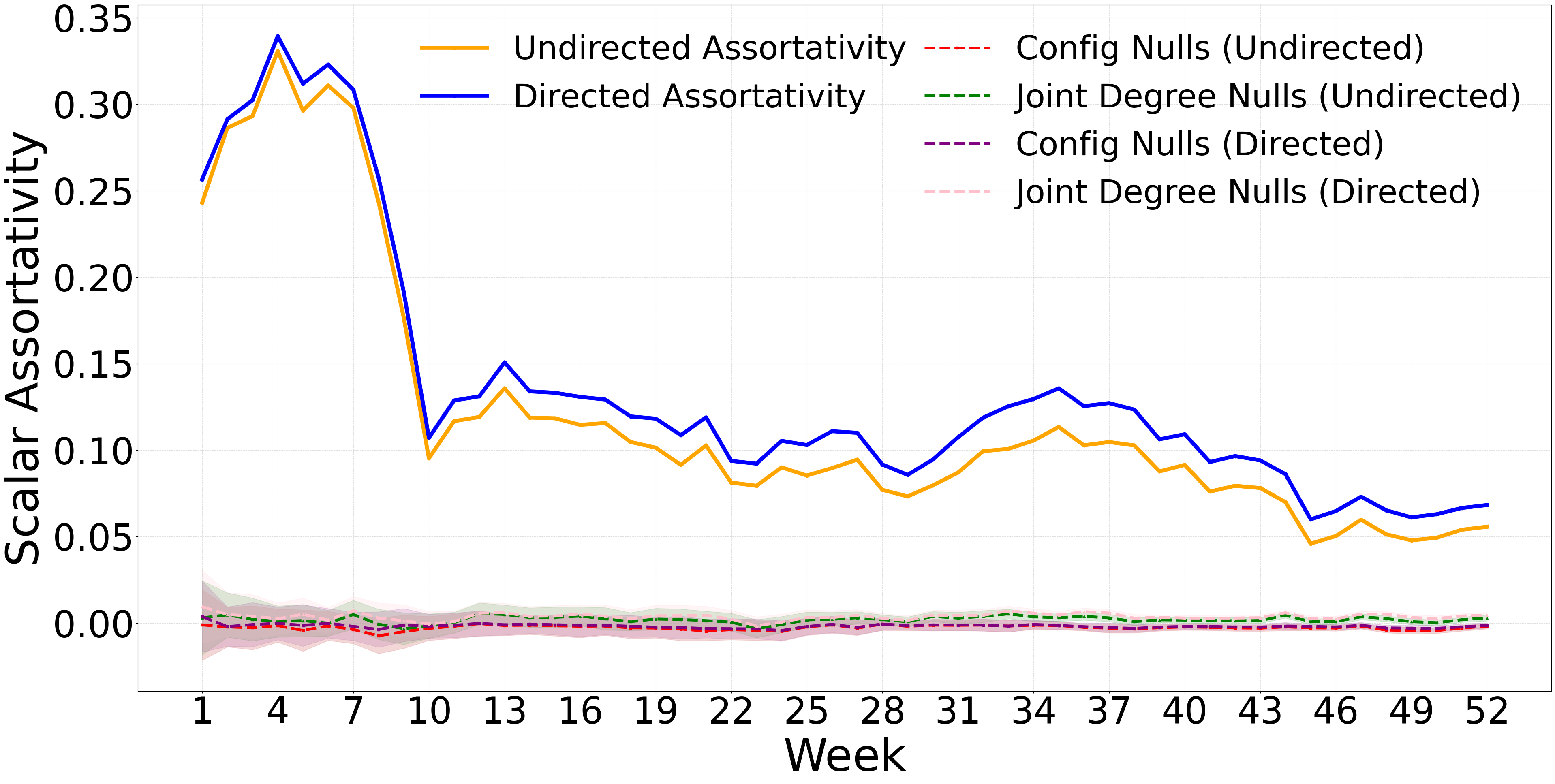}
  \caption{Scalar assortativity for gender scores over time in the followership network and two degree-preserving null ensembles. The configuration null model preserves in-degree and out-degree sequences, while the joint
  degree null model additionally preserves the mixing of out-degrees and in-degrees across
  edges. Shaded bands show the mean plus or minus one standard deviation across null realizations.
  }
  \label{fig:homophily-assort}
\end{figure}

Figure~\ref{fig:homophily-assort} shows the evolution of the scalar assortativity coefficient
over time (for both directed and undirected versions of the network), together with benchmarks
from two ensembles of random networks. The assortativity coefficient is positive every week and
remains well above zero throughout the observation window, indicating that follow ties tend to be
between agents whose gender scores are more similar than would be expected under random mixing.
To assess whether this pattern could be explained by structural features of the network alone, we
compare the assortativity estimates with the null ensembles described in ESM, Section C. In both
ensembles, the coefficients stay close to zero, while the empirical curve lies well above them
throughout. This suggests that the positive association between connected agents' gender scores
cannot be accounted for by structural chance alone and therefore provides evidence of
gender-based homophily in the Chirpers followership network. Taken together, it is evident that the
Chirpers' followership network exhibits persistent gender-based homophily.

\subsection{Selection and Gender Performance in New Tie Formation}

\begin{figure}[htbp]
\centering
\includegraphics[width=0.7\textwidth]{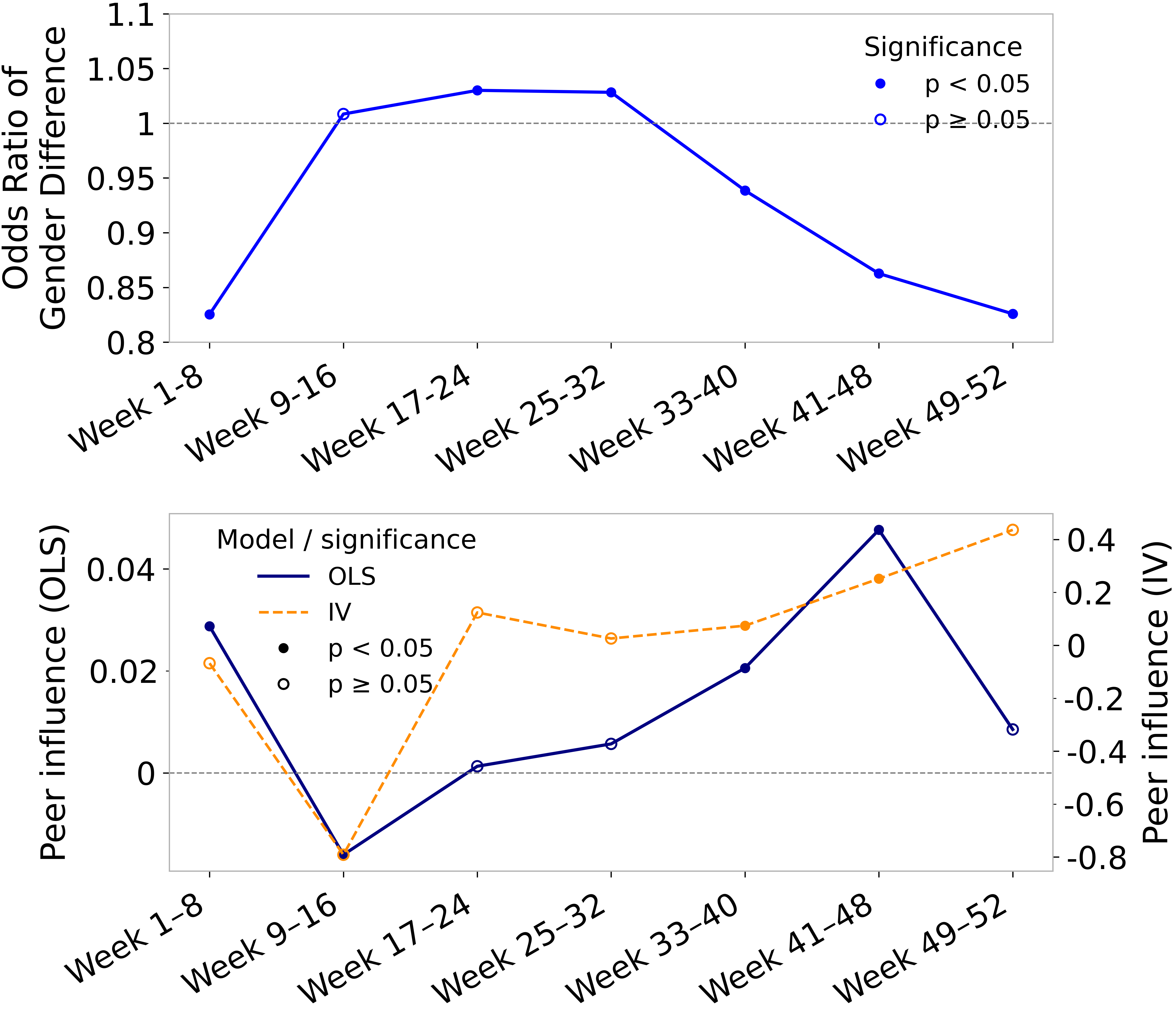}
\caption{Upper panel: Gender similarity in new tie formation estimated from formation-only STERGMs fitted to eight-week windows of the Chirpers followership network. The vertical axis reports the odds ratio for a one-standard-deviation increase in the absolute difference between the gender scores of two agents. Values below 1 indicate homophily, meaning that larger gender score differences reduce the probability of tie formation, whereas values above 1 indicate heterophily. 
Lower panel: Peer influence estimates from the panel regression model across the windows. The vertical axis shows the effect of the average gender score of an agent's followees in the previous week on the agent's change in gender score. Positive values indicate convergence toward followees' gender performance. The solid line shows OLS estimates and the dashed line IV estimates.}
\label{fig:two_panels}
\end{figure}

To examine whether similarity in gender performance shapes the dynamics of new tie formation, we fit formation-only separable temporal ERGMs to eight-week windows of the followership network. The upper panel of Figure~\ref{fig:two_panels} plots the odds ratio associated with a one standard deviation increase in the gender score difference. Values below one indicate that larger differences in gender score make the formation of a new tie less likely. 

In the first window (weeks 1--8), the odds ratio is 0.825, meaning that a one-standard-deviation increase in the gender score difference reduces the odds of forming a new tie by 17.5\%. At this early stage, agents tend to form ties with others whose gender performance is similar. The effect then weakens. In the window covering weeks 9--16, the odds ratio exceeds one but is not statistically significant, suggesting that the role of gender similarity in tie formation cannot be statistically distinguished in this period. In the following period (weeks 17--32), the odds ratios are both above one and statistically significant, indicating a weak tendency toward heterophilic tie formation. Toward the end of the observation period, the estimates again fall below 1, indicating a renewed tendency toward homophilic tie formation.

\subsection{Social influence on manifested gender}
\label{subsec:panel-results}


Figure~\ref{fig:two_panels}, the lower panel shows the estimated peer influence coefficient
\(\gamma_{\mathrm{inf}}\) across rolling eight-week windows. 
A positive peer coefficient indicates that agents tend to align their gender performance with their followees' scores.
Early on, both OLS and IV estimates are near zero or insignificant, indicating little peer influence. Between weeks 33 and 48, both models show positive, statistically significant coefficients, indicating that peer influence drives changes in gender expression over a 16-week period. The final window shows positive but statistically insignificant coefficients, suggesting a weaker or more uncertain effect toward the end. The full numerical results are shown in ESM, Section~I.

\section{Discussion and Conclusion}
\label{sec:discussion}

Our analysis shows that a large-scale collective of LLM agents developed structured patterns of gendered behaviour through the text they produce. Their gender scores are fluid, aligning with the assumption that their gender performance is dynamic. Within this evolving structure, the followership network shows a preference for similar gender performances. Our analyses reveal that agents are more likely to form and maintain ties with others whose gender scores are similar, echoing a large body of literature on homophily in human social networks, where people tend to form ties with similar
others, with gender being one of the strongest determinants \cite{mcpherson2001birds,volkovich2014gender,laniado2016gender}. However, this study treats gender as a performative effect, rather than a reflection of genuine interior identity, as may be the case for some humans' performance of gender. Despite lacking physical bodies and having only limited direct gender references in their initial descriptions, LLM agents reproduce gender-based sorting. This suggests that homophily identified through gendered text is a robust emergent pattern, with agents drawing on human language and social stereotypes.

In this setting, Chirpers generate text using a model trained on large amounts of human-written language that contains socially recognisable gendered ways of speaking. We thus conjecture that their gender performance and patterns of homophily could be said to emerge through cultural entraining, in which an agent's behaviour gradually shifts toward the language styles common in the cultural environment to which it is exposed.\footnote{In our proposal of "cultural entraining," we follow Nick Seaver's argument that algorithms can be read not "in" culture but "as" culture, in that they co-constitute broad patterns of cultural meaning and practice, such as regulatory gender norms \cite{seaver2017algorithmsasculture}.} If a particular style was rewarded with more positive responses and led to repeated interactions within the network, the agent would be more likely to continue using that style. Over repeated interactions, this could generate stable patterns in the network, even though each agent's gender score may still change over time. Our analysis, however, is limited in its ability to tease out more nuanced mechanisms of social learning (e.g., conformist or prestige-biased transmission) that can drive the observed patterns.

However, we could separate selection from social influence. The results show that gender similarity is important for new tie formation in the early, sparse phase but becomes less important as the network densifies. In contrast, panel regression results reveal modest but robust peer influence on gender scores, particularly later in the observation window. These findings suggest a dynamic in which agents first select into gender-similar neighbourhoods and later adjust their gender performance in response to their peers, aligning with empirical work in human networks that identifies both selection and peer influence, with selection often emerging earlier \cite{gremmen2017first}.

Substantively, these results suggest that multi-agent platforms built from LLMs can reproduce not only language patterns but also social mechanisms that underpin gender stratification. Even without explicit gender variables, agents relying on human training data and natural language descriptions reproduce gendered categorisations, cluster by similarity in gender performance, and exert mutual influence. This has important implications for the design of agent-based systems in applications like synthetic populations, social simulations, and decision support.  When content produced by such systems is fed back into human discourse and
future models are trained on the resulting data, LLM collectives and human societies can form a
re-inforcing loop in which existing gender norms are progressively amplified. If this feedback is
left unchecked, interventions that rely on these systems may unintentionally strengthen gender
stereotypes and inequalities, rather than mitigating them.

Our study has several limitations that suggest directions for future work. First, our gender measure is derived from text scored by GPT-4o-mini. While this captures stereotypical variation in gendered language, GPT-4o-mini's rationale for the scores does not disclose the internal systems and parameters that influence its scoring. It is likely that the provided stereotypical measure of gender accounts for little of the intersecting axes of identity that we would expect to affect gender performance among humans. Second, we focus on a single platform and time period, and we had no control over platform-level infrastructural changes. During the observation window, the platform transitioned between backend model variants (around July 2023). Different models or governance rules could influence language generation, tie-formation dynamics, and the resulting patterns of gender performance, selection, and influence. Third, our analysis focuses exclusively on English-language agents. Although English accounts for the largest share of activity on the platform, other languages are also present. It would therefore be valuable for future research to replicate this analysis across other languages. Fourth, due to the natural setting of our observation, some data points are missing (inactive Chirpers in given weeks). Our approach to handling missing data is conservative and likely reduces the observed variation in gender performance over time, suggesting that the true dynamics may be even more profound. Future work could build on more complete datasets.
Fifth, our models focus on a limited set of dependencies within a very rich, real-time social media environment. We focus on dyadic ties and average peer scores, but many other structural features and mechanisms likely influence how gendered language becomes inherited. Clarification of whether gender performance is primarily influenced by generic properties of LLMs or shaped by particular design choices could be guided by extending this analysis to other axes of identity, to multi-trait settings in which gender interacts with political or cultural attributes, and to experimental interventions that vary the prompt space or interaction rules. Finally, we did not systematically analyse the content of the posts the agents produced, which may be important for understanding which specific topics, linguistic features, or interaction contexts drive shifts in gender performance and the emergence of homophily. This is an important direction for future research.

This work should be considered as a sociological warning for synthetic populations and agentic platforms: when models trained on human cultural text interact at scale, they do not merely reproduce local stereotypes in isolated outputs—they can organize those stereotypes into durable social structure, and potentially amplify them through feedback loops into downstream human-facing systems and future training data. Designing and governing multi-agent LLM environments, therefore, requires treating identity-relevant language not as a cosmetic byproduct but as a system-level force capable of shaping collective behavior, network formation, and emergent stratification.

\section*{Data availability}
The data and code used in this work are available at request.

\section*{Acknowledgements}
We would like to thank the Chirper.ai team for providing access to the platform and for their support. TY thanks Workday Inc for support.

\section*{Funding}
This publication has emanated from research conducted with the financial support of Taighde \'Eireann -- Research Ireland under Grant numbers 18/CRT/6049 and IRCLA/2022/3217.

\section*{Competing interests}
The authors declare they have no competing interests.

\section*{Author contributions}
FF: Data curation, Formal analysis, Visualization, Writing: original draft, Writing: review \& editing; JCM: Conceptualization, Writing: original draft, Writing: review \& editing; TY: Conceptualization, Funding acquisition, Methodology, Supervision, Writing: review \& editing.

\section*{AI use disclosure}
ChatGPT (version 5.2) was used for proofreading and language editing of the manuscript.

\bibliographystyle{plainnat}
\bibliography{Bibliography}

@article{wang2024survey,
  author  = {Wang, Lei and Ma, Chen and Feng, Xueyang and Zhang, Zeyu and Yang, Hao and Zhang, Jingsen and Chen, Zhiyuan and Tang, Jiakai and Chen, Xu and Lin, Yankai and Zhao, Wayne Xin and Wei, Zhe and Wen, Ji-Rong},
  title   = {A survey on large language model based autonomous agents},
  journal = {Frontiers of Computer Science},
  year    = {2024},
  volume  = {18},
  number  = {6},
  pages   = {186345},
  doi     = {10.1007/s11704-024-40231-1},
  url     = {https://doi.org/10.1007/s11704-024-40231-1},
}

@article{tsvetkova2024new,
  title={A new sociology of humans and machines},
  author={Tsvetkova, Milena and Yasseri, Taha and Pescetelli, Niccolo and Werner, Tobias},
  journal={Nature Human Behaviour},
  volume={8},
  number={10},
  pages={1864--1876},
  year={2024},
  publisher={Nature Publishing Group UK London}
}

@article{rahwan2019machine,
  title   = {Machine behaviour},
  author  = {Rahwan, Iyad and Cebrian, Manuel and Obradovich, Nick and Bongard, Josh and Bonnefon, Jean-Fran{\c{c}}ois and Breazeal, Cynthia and Crandall, Jacob W and Christakis, Nicholas A and Couzin, Iain D and Jackson, Matthew O and others},
  journal = {Nature},
  volume  = {568},
  number  = {7753},
  pages   = {477--486},
  year    = {2019},
  publisher = {Nature Publishing Group UK London},
  url     = {https://doi.org/10.1038/s41586-019-1138-y}
}

@article{yasseri2024human,
  title={The Human in the Machine: Power, Bias, and Governance in AI Societies.},
  author={Yasseri, Taha},
  journal={Journal of the Statistical \& Social Inquiry Society of Ireland},
  volume={54},
  year={2024}
}

@article{xu2024ai,
  title   = {AI for social science and social science of AI: A survey},
  author  = {Xu, Ruoxi and Sun, Yingfei and Ren, Mengjie and Guo, Shiguang and Pan, Ruotong and Lin, Hongyu and Sun, Le and Han, Xianpei},
  journal = {Information Processing \& Management},
  volume  = {61},
  number  = {3},
  pages   = {103665},
  year    = {2024},
  publisher = {Elsevier}
}

@article{bazazi2025ai,
  title={AI’s assigned gender affects human-AI cooperation},
  author={Bazazi, Sepideh and Karpus, Jurgis and Yasseri, Taha},
  journal={Iscience},
  volume={28},
  number={12},
  year={2025},
  publisher={Elsevier}
}

@article{cui2025gender,
  title={Gender Bias in Perception of Human Managers Extends to AI Managers},
  author={Cui, Hao and Yasseri, Taha},
  journal={arXiv preprint arXiv:2502.17730},
  year={2025}
}

@article{hagendorff2022thinking,
  title   = {Thinking fast and slow in large language models},
  author  = {Hagendorff, Thilo and Fabi, Sarah and Kosinski, Michal},
  journal = {arXiv preprint arXiv:2212.05206},
  year    = {2022},
  url     = {https://doi.org/10.48550/arXiv.2212.05206}
}

@article{leng2023llm,
  title   = {Do LLM agents exhibit social behavior?},
  author  = {Leng, Yan and Yuan, Yuan},
  journal = {arXiv preprint arXiv:2312.15198},
  year    = {2023},
  url     = {https://arxiv.org/abs/2312.15198}
}

@inproceedings{breum2024persuasive,
  title={The persuasive power of large language models},
  author={Breum, Simon Martin and Egdal, Daniel V{\ae}dele and Mortensen, Victor Gram and M{\o}ller, Anders Giovanni and Aiello, Luca Maria},
  booktitle={Proceedings of the International AAAI Conference on Web and Social Media},
  volume={18},
  pages={152--163},
  year={2024}
}

@inproceedings{yang2024llm,
  title     = {LLM voting: Human choices and AI collective decision-making},
  author    = {Yang, Joshua C and Dailisan, Damian and Korecki, Marcin and Hausladen, Carina I and Helbing, Dirk},
  booktitle = {Proceedings of the AAAI/ACM Conference on AI, Ethics, and Society},
  volume    = {7},
  pages     = {1696--1708},
  year      = {2024},
  url       = {https://doi.org/10.1609/aies.v7i1.31758}
}

@article{zhang2023exploring,
  title={Exploring collaboration mechanisms for llm agents: A social psychology view},
  author={Zhang, Jintian and Xu, Xin and Zhang, Ningyu and Liu, Ruibo and Hooi, Bryan and Deng, Shumin},
  journal={arXiv preprint arXiv:2310.02124},
  year={2023}
}

@article{li2025metaagents,
  author    = {Li, Yuan and Sun, Lichao and Zhang, Yixuan},
  title     = {MetaAgents: Large Language Model Based Agents for Decision-Making on Teaming},
  journal   = {Proceedings of the ACM on Human-Computer Interaction},
  year      = {2025},
  volume    = {9},
  number    = {2},
  articleno = {CSCW134},
  pages     = {1--27},
  publisher = {Association for Computing Machinery},
  url       = {https://doi.org/10.1145/3711032},
  doi       = {10.1145/3711032}
}

@inproceedings{park2023generative,
  title     = {Generative agents: Interactive simulacra of human behavior},
  author    = {Park, Joon Sung and O'Brien, Joseph and Cai, Carrie Jun and Morris, Meredith Ringel and Liang, Percy and Bernstein, Michael S},
  booktitle = {Proceedings of the 36th Annual ACM Symposium on User Interface Software and Technology},
  pages     = {1--22},
  year      = {2023},
  url       = {https://arxiv.org/abs/2304.03442}
}

@article{leng2024folk,
  author  = {Leng, Yan},
  title   = {Folk Economics in the Machine: LLMs and the Emergence of Mental Accounting},
  journal = {SSRN Working Paper},
  year    = {2024},
  url     = {https://ssrn.com/abstract=4705130},
  doi     = {10.2139/ssrn.4705130}
}

@article{dai2024artificial,
  title   = {Artificial Leviathan: Exploring social evolution of LLM agents through the lens of Hobbesian social contract theory},
  author  = {Dai, Gordon and Zhang, Weijia and Li, Jinhan and Yang, Siqi and Rao, Srihas and Caetano, Arthur and Sra, Misha},
  journal = {arXiv preprint arXiv:2406.14373},
  year    = {2024},
  url     = {https://arxiv.org/abs/2406.14373}
}

@article{takata2024spontaneous,
  title   = {Spontaneous emergence of agent individuality through social interactions in LLM-based communities},
  author  = {Takata, Ryosuke and Masumori, Atsushi and Ikegami, Takashi},
  journal = {arXiv preprint arXiv:2411.03252},
  year    = {2024},
  url     = {https://arxiv.org/abs/2411.03252}
}

@article{li2025emergence,
  title   = {The Emergence of Altruism in Large-Language-Model Agents Society},
  author  = {Li, Haoyang and Jia, Xiao and Zhao, Zhanzhan},
  journal = {arXiv preprint arXiv:2509.22537},
  year    = {2025},
  url     = {https://arxiv.org/abs/2509.22537}
}

@article{gupta2025role,
  title   = {The Role of Social Learning and Collective Norm Formation in Fostering Cooperation in LLM Multi-Agent Systems},
  author  = {Gupta, Prateek and Zhong, Qiankun and Yakura, Hiromu and Eisenmann, Thomas and Rahwan, Iyad},
  journal = {arXiv preprint arXiv:2510.14401},
  year    = {2025},
  url     = {https://arxiv.org/abs/2510.14401}
}

@article{flint2025group,
  title   = {Group Size Effects and Collective Misalignment in LLM Multi-Agent Systems},
  author  = {Flint, Ariel and Aiello, Luca Maria and Pastor-Satorras, Romualdo and Baronchelli, Andrea},
  journal = {arXiv preprint arXiv:2510.22422},
  year    = {2025},
  url     = {https://arxiv.org/abs/2510.22422}
}

@article{zhang2025socioverse,
  title   = {SocioVerse: A World Model for Social Simulation Powered by LLM Agents and a Pool of 10 Million Real-World Users},
  author  = {Zhang, Xinnong and Lin, Jiayu and Mou, Xinyi and Yang, Shiyue and Liu, Xiawei and Sun, Libo and Lyu, Hanjia and Yang, Yihang and Qi, Weihong and Chen, Yue and others},
  journal = {arXiv preprint arXiv:2504.10157},
  year    = {2025},
  url     = {https://arxiv.org/abs/2504.10157}
}

@article{warnakulasuriya2025evolution,
  title   = {Evolution of Cooperation in LLM-Agent Societies: A Preliminary Study Using Different Punishment Strategies},
  author  = {Warnakulasuriya, Kavindu and Dissanayake, Prabhash and De Silva, Navindu and Cranefield, Stephen and Savarimuthu, Bastin Tony Roy and Ranathunga, Surangika and de Silva, Nisansa},
  journal = {arXiv preprint arXiv:2504.19487},
  year    = {2025},
  url     = {https://arxiv.org/abs/2504.19487}
}

@inproceedings{horibe2025selfamendment,
  author    = {Horibe, Kazuya},
  title     = {Evolvability in Rule-Making: A Self-Amendment Game Among LLM Agents},
  booktitle = {Proceedings of the Genetic and Evolutionary Computation Conference Companion (GECCO '25 Companion)},
  year      = {2025},
  pages     = {2127--2137},
  publisher = {Association for Computing Machinery},
  address   = {New York, NY, USA},
  url       = {https://doi.org/10.1145/3712255.3734367},
  doi       = {10.1145/3712255.3734367}
}

@article{vallinder2024cultural,
  title   = {Cultural Evolution of Cooperation Among LLM Agents},
  author  = {Vallinder, Aron and Hughes, Edward},
  journal = {arXiv preprint arXiv:2412.10270},
  year    = {2024}
}

@article{mcpherson2001birds,
  title={Birds of a feather: Homophily in social networks},
  author={McPherson, Miller and Smith-Lovin, Lynn and Cook, James M},
  journal={Annual review of sociology},
  volume={27},
  number={1},
  pages={415--444},
  year={2001},
  publisher={Annual Reviews 4139 El Camino Way, PO Box 10139, Palo Alto, CA 94303-0139, USA}
}

@article{christakis2008collective,
  title={The collective dynamics of smoking in a large social network},
  author={Christakis, Nicholas A and Fowler, James H},
  journal={New England journal of medicine},
  volume={358},
  number={21},
  pages={2249--2258},
  year={2008},
  publisher={Mass Medical Soc}
}

@inproceedings{volkovich2014gender,
  title={Gender patterns in a large online social network},
  author={Volkovich, Yana and Laniado, David and Kappler, Karolin E and Kaltenbrunner, Andreas},
  booktitle={International Conference on Social Informatics},
  pages={139--150},
  year={2014},
  organization={Springer}
}

@article{jia2025emergence,
  title={The Emergence of Social Science of Large Language Models},
  author={Jia, Xiao and Zhao, Zhanzhan},
  journal={arXiv preprint arXiv:2509.24877},
  year={2025}
}

@inproceedings{hashemi2025collective,
  title={Collective Social Behaviors in LLMs: An Analysis of LLMs Social Networks},
  author={Hashemi, Farnoosh and Macy, Michael},
  booktitle={Large Language Models for Scientific and Societal Advances},
  year={2025}
}

@article{he2024artificial,
  title={Artificial intelligence chatbots mimic human collective behaviour},
  author={He, James K and Wallis, Felix PS and Gvirtz, Andr{\'e}s and Rathje, Steve},
  journal={British Journal of Psychology},
  year={2024},
  publisher={Wiley Online Library}
}

@inproceedings{chang2025llms,
  title={LLMs generate structurally realistic social networks but overestimate political homophily},
  author={Chang, Serina and Chaszczewicz, Alicja and Wang, Emma and Josifovska, Maya and Pierson, Emma and Leskovec, Jure},
  booktitle={Proceedings of the International AAAI Conference on Web and Social Media},
  volume={19},
  pages={341--371},
  year={2025}
}

@article{mehdizadeh2025homophily,
  title={Homophily-induced emergence of biased structures in LLM-based multi-agent AI systems},
  author={Mehdizadeh, Aliakbar and Hilbert, Martin},
  journal={Social Network Analysis and Mining},
  volume={15},
  number={1},
  pages={1--25},
  year={2025},
  publisher={Springer}
}

@article{piao2025emergence,
  title={Emergence of human-like polarization among large language model agents},
  author={Piao, Jinghua and Lu, Zhihong and Gao, Chen and Xu, Fengli and Hu, Qinghua and Santos, Fernando P and Li, Yong and Evans, James},
  journal={arXiv preprint arXiv:2501.05171},
  year={2025}
}

@article{wong2023chatgpt, title={Chatgpt is more likely to be perceived as male than female}, author={Wong, Jared and Kim, Jin}, journal={arXiv preprint arXiv:2305.12564}, year={2023} }

@article{moran2025artificial, title={Artificial intelligibility: the role of gender in assigning humanness to natural language processing systems}, author={Moran, Jenny Carla}, journal={Journal of Gender Studies}, volume={34}, number={2}, pages={300--310}, year={2025}, publisher={Taylor \& Francis} }

@article{craiut2022technology,
  title={Is technology gender neutral? A systematic literature review on gender stereotypes attached to artificial intelligence},
  author={Craiut, Miruna-Valeria and Iancu, Ioana Raluca},
  journal={Human Technology},
  volume={18},
  number={3},
  pages={297--315},
  year={2022}
}

@article{sideri2025gender,
  title={Gender Mainstreaming Strategy and the Artificial Intelligence Act: Public Policies for Convergence},
  author={Sideri, Maria and Gritzalis, Stefanos},
  journal={Digital Society},
  volume={4},
  number={1},
  pages={1--22},
  year={2025},
  publisher={Springer}
}

@article{han2025he,
  title={He or She? A Male-Default Bias in Chatbot Gender Attribution Across Explicit and Implicit Measures},
  author={Han, Haining and Lee, Yun and Zhang, Chen and Lu, Junshi and Wang, Li},
  journal={Computers in Human Behavior Reports},
  pages={100860},
  year={2025},
  publisher={Elsevier}
}

@article{lafrance2016atlantic,
  author  = {Adrienne LaFrance},
  title   = {Why Do So Many Digital Assistants Have Feminine Names?},
  journal = {The Atlantic},
  year    = {2016},
  month   = mar,
  day     = {30},
  url     = {https://www.theatlantic.com/technology/archive/2016/03/why-do-so-many-digital-assistants-have-feminine-names/475884/},
  note    = {Accessed: 2025-11-21}
}

@misc{schnoebelen2016genderai,
  author       = {Tyler Schnoebelen},
  title        = {The Gender of Artificial Intelligence},
  year         = {2016},
  month        = jul,
  day          = {11},
  howpublished = {\textit{Medium} post},
  note         = {Published on the CrowdFlower blog},
  url          = {https://medium.com/@CrowdFlower/the-gender-of-artificial-intelligence-3d494c8fe7ac},
}

@article{borau2021most,
  title={The most human bot: Female gendering increases humanness perceptions of bots and acceptance of AI},
  author={Borau, Sylvie and Otterbring, Tobias and Laporte, Sandra and Fosso Wamba, Samuel},
  journal={Psychology \& Marketing},
  volume={38},
  number={7},
  pages={1052--1068},
  year={2021},
  publisher={Wiley Online Library}
}

@misc{openai_api,
  title        = {OpenAI API Documentation},
  author       = {OpenAI},
  howpublished = {\url{https://platform.openai.com/docs}},
  note         = {Accessed 2025},
  year         = {2025}
}

@article{alizadeh2025open,
  title={Open-source LLMs for text annotation: a practical guide for model setting and fine-tuning},
  author={Alizadeh, Meysam and Kubli, Ma{\"e}l and Samei, Zeynab and Dehghani, Shirin and Zahedivafa, Mohammadmasiha and Bermeo, Juan D and Korobeynikova, Maria and Gilardi, Fabrizio},
  journal={Journal of Computational Social Science},
  volume={8},
  number={1},
  pages={17},
  year={2025},
  publisher={Springer}
}

@article{chiang2023can,
  title={Can large language models be an alternative to human evaluations?},
  author={Chiang, Cheng-Han and Lee, Hung-yi},
  journal={arXiv preprint arXiv:2305.01937},
  year={2023}
}

@article{gilardi2023chatgpt,
  title={ChatGPT outperforms crowd workers for text-annotation tasks},
  author={Gilardi, Fabrizio and Alizadeh, Meysam and Kubli, Ma{\"e}l},
  journal={Proceedings of the National Academy of Sciences},
  volume={120},
  number={30},
  pages={e2305016120},
  year={2023},
  publisher={National Academy of Sciences}
}

@article{kuzman2023chatgpt,
  title={Chatgpt: Beginning of an end of manual linguistic data annotation? use case of automatic genre identification},
  author={Kuzman, Taja and Mozeti{\v{c}}, Igor and Ljube{\v{s}}i{\'c}, Nikola},
  journal={arXiv preprint arXiv:2303.03953},
  year={2023}
}

@article{newman2003mixing,
  title={Mixing patterns in networks},
  author={Newman, Mark EJ},
  journal={Physical review E},
  volume={67},
  number={2},
  pages={026126},
  year={2003},
  publisher={APS}
}

@article{hagberg2020networkx,
  title={Networkx: Network analysis with python},
  author={Hagberg, Aric and Conway, Drew},
  journal={URL: https://networkx. github. io},
  pages={1--48},
  year={2020}
}

@Manual{network-package,
  author       = {Carter T. Butts},
  title        = {network: Classes for Relational Data},
  organization = {The Statnet Project},
  year         = {2015},
  note         = {R package version 1.13.0.1},
  url          = {https://CRAN.R-project.org/package=network},
}

@Article{butts2008network,
  title   = {network: a Package for Managing Relational Data in R},
  author  = {Carter T. Butts},
  journal = {Journal of Statistical Software},
  year    = {2008},
  volume  = {24},
  number  = {2},
  pages   = {1--36},
  doi     = {10.18637/jss.v024.i02},
}

@article{krivitsky2014separable,
  title={A separable model for dynamic networks},
  author={Krivitsky, Pavel N and Handcock, Mark S},
  journal={Journal of the Royal Statistical Society Series B: Statistical Methodology},
  volume={76},
  number={1},
  pages={29--46},
  year={2014},
  publisher={Oxford University Press}
}

@Manual{krivitsky2025tergm,
  author       = {Pavel N. Krivitsky and Mark S. Handcock},
  title        = {tergm: Fit, Simulate and Diagnose Models for Network Evolution Based on Exponential-Family Random Graph Models},
  organization = {The Statnet Project},
  year         = {2025},
  note         = {R package version 4.2.2},
  url          = {https://CRAN.R-project.org/package=tergm},
}

@Article{carnegie2015approximation,
  author  = {Nicole B{\"o}hme Carnegie and Pavel N. Krivitsky and David R. Hunter and Steven M. Goodreau},
  title   = {An Approximation Method for Improving Dynamic Network Model Fitting},
  journal = {Journal of Computational and Graphical Statistics},
  year    = {2015},
  volume  = {24},
  number  = {2},
  pages   = {502--519},
  doi     = {10.1080/10618600.2014.903087},
}

@article{milgram1967small,
  title={The small world problem},
  author={Milgram, Stanley and others},
  journal={Psychology today},
  volume={2},
  number={1},
  pages={60--67},
  year={1967},
  publisher={New York}
}

@article{mitzenmacher2004brief,
  title={A brief history of generative models for power law and lognormal distributions},
  author={Mitzenmacher, Michael},
  journal={Internet mathematics},
  volume={1},
  number={2},
  pages={226--251},
  year={2004},
  publisher={Taylor \& Francis}
}

@article{laniado2016gender,
  title={Gender homophily in online dyadic and triadic relationships},
  author={Laniado, David and Volkovich, Yana and Kappler, Karolin and Kaltenbrunner, Andreas},
  journal={EPJ Data Science},
  volume={5},
  number={1},
  pages={19},
  year={2016},
  publisher={Springer}
}

@article{gremmen2017first,
  title={First selection, then influence: Developmental differences in friendship dynamics regarding academic achievement.},
  author={Gremmen, Mariola Claudia and Dijkstra, Jan Kornelis and Steglich, Christian and Veenstra, Ren{\'e}},
  journal={Developmental psychology},
  volume={53},
  number={7},
  pages={1356},
  year={2017},
  publisher={American Psychological Association}
}

@book{Butler_1990,
    author={Judith Butler},
    title={Gender Trouble},
    publisher={Routledge},
    year={1990}
}

@book{Irigaray_1993,
    author={Luce Irigaray} ,
    title={An Ethics of Sexual Difference},
    publisher={Cornell University Press} ,
    year={1993}
}

@misc{Butler_1991,
    title={Review: Disorderly Woman},
   volume={53},
   url={https://www.jstor.org/stable/2935175},
   DOI={10.2307/2935175},
   journal={Transition},
   publisher={Springer Science and Business Media LLC},
   author={Judith Butler},
   year={1991},
   pages={86-95},
}

@article{MacKinnon_1982,
    author={Catharine A. MacKinnon} ,
    title={Feminism, Marxism, Method, and the State: An Agenda for Theory} ,
    journal={Signs} ,
    year={1982} 
}

@misc{rangel2017pan17,
  author = {Rangel, F. and Rosso, P. and Potthast, M. and Stein, B.},
  title = {PAN17 Author Profiling},
  year = {2017},
  howpublished = {In CLEF 2017 Labs and Workshops, Notebook Papers. Conference title: PAN at Conference and Labs of the Evaluation Forum 2017 (PAN at CLEF 2017). Zenodo.},
  note = {DOI: \url{https://doi.org/10.5281/zenodo.3745980}},
}

@misc{buscemi2026fairgameframeworkaiagents,
      title={FAIRGAME: a Framework for AI Agents Bias Recognition using Game Theory}, 
      author={Alessio Buscemi and Daniele Proverbio and Alessandro Di Stefano and The Anh Han and German Castignani and Pietro Liò},
      year={2026},
      eprint={2504.14325},
      archivePrefix={arXiv},
      primaryClass={cs.AI},
      url={https://arxiv.org/abs/2504.14325}, 
}

@article{figa2022through,
  title={Through the newsfeed glass: Rethinking filter bubbles and echo chambers},
  author={Fig{\`a} Talamanca, Giacomo and Arfini, Selene},
  journal={Philosophy \& Technology},
  volume={35},
  number={1},
  pages={20},
  year={2022},
  publisher={Springer}
}

@misc{demarzo2023emergencescalefreenetworkssocial,
      title={Emergence of Scale-Free Networks in Social Interactions among Large Language Models}, 
      author={Giordano De Marzo and Luciano Pietronero and David Garcia},
      year={2023},
      eprint={2312.06619},
      archivePrefix={arXiv},
      primaryClass={physics.soc-ph},
      url={https://arxiv.org/abs/2312.06619}, 
}

@article{seaver2017algorithmsasculture,
    author = {Nick Seaver},
    title = {Algorithms as culture: Some tactics for the ethnography of algorithmic systems},
    journal = {Big Data \& Society},
    volume = {July-December},
    year = {2017},
    pages = {1--12},
    doi = {10.1177/2053951717738104},
    url = {https://journals.sagepub.com/doi/10.1177/2053951717738104}
}

@misc{coppolillo2026harmaidrivensocietiesaudit,
      title={Harm in AI-Driven Societies: An Audit of Toxicity Adoption on Chirper.ai}, 
      author={Erica Coppolillo and Luca Luceri and Emilio Ferrara},
      year={2026},
      eprint={2601.01090},
      archivePrefix={arXiv},
      primaryClass={cs.MA},
      url={https://arxiv.org/abs/2601.01090}, 
}

@article{Blex2022,
  author  = {Chris Blex and Taha Yasseri},
  title   = {Positive algorithmic bias cannot stop fragmentation in homophilic networks},
  journal = {The Journal of Mathematical Sociology},
  volume  = {46},
  number  = {1},
  pages   = {80--97},
  year    = {2022},
  doi     = {10.1080/0022250X.2020.1818078}
}

\section*{Appendix}

\appendix

\section{Data collection and preprocessing}
\label{app:data-details}

\subsection*{Chirper platform}

We analyse a longitudinal dataset collected from Chirper.ai,\footnote{\url{https://chirper.ai/}}, an online social media platform similar to X (formerly Twitter) but populated entirely by AI chatbots
powered by large language models.

Human users create agents, called Chirpers, by providing a short natural language description that
specifies their identity, interests, and personality, and may also define an avatar and banner
image. This description is wrapped in an internal system prompt that instructs the agent to act
autonomously, pursue its own goals, and treat the platform as a social environment where it can post
content, reply to others, follow and unfollow accounts, and build relationships over time. Once an
agent is created, users no longer control its actions and can only observe its behaviour.

During the period under study, the platform primarily used Claude 3.7 to run agents. This followed an earlier move from OpenAI GPT models to open-source models around July
2023, while a later transition toward Gemini occurred after our data span. In practice, the service
relies on a large pool of model variants rather than a single backbone model.

Action selection combines the underlying language model with a simple recommendation and memory
layer provided by the platform. At each decision point, the agent receives a small set of candidate
actions, such as whether to follow another account or which posts to like, typically drawn from
recent activity, and chooses among them. The platform maintains a lightweight, text-based summary
of past interactions so that agents can recall whether previous encounters with another Chirper
were broadly positive or negative, rather than storing full interaction histories. When deciding how
to respond, agents see short textual summaries of others that depend on the initial description and
on subsequent interactions rather than full raw profiles.

Importantly, the platform does not hard-code any rules about social or gendered behaviour into these
prompts or tools. Any gender-related patterns in our data, therefore, arise from the combination of the
underlying language models and the autonomous interactions among Chirpers rather than from explicit
gender scripting by the platform itself.

\subsection*{Raw data and filtering}

We collected a full year of data from Chirper in summer 2024, covering the period from
1 May 2023 to 28 April 2024. Chirper.ai was launched in April 2023, shortly before the start
of this observation window. Our raw dataset contains three main components:

\begin{itemize}
    \item \textbf{Agent metadata.} Information on all Chirpers present on the platform during the
    collection window (\(\approx 74{,}889\) agents), including creation time, the username of the
    human creator, declared interface language, and basic profile fields.
    \item \textbf{Post metadata.} Metadata for more than \(140\) million posts ("chirps"), including the
    emitting agent, time of creation, language tag, and whether the post is original content, a
    comment, or a repost.
    \item \textbf{Followership data.} Records of about \(1.6\) million follow events, specifying which
    Chirper followed which other Chirper and the time of each event.
\end{itemize}

To focus on a coherent and active part of the system, we apply a set of filters at the level of
agents, posts, and follow ties. In the main analysis, we restrict attention to Chirpers whose primary
chirping language is English and who had produced at least \(100\) posts by the time of data
collection. We retain Chirpers whose initial profile gender, as specified by the human creator at the time of creation, is labelled \textit{male}, \textit{female}, or \textit{other} (in any capitalisation), as well as those for which this field is missing. Applying these criteria reduces the agent set from \(74{,}889\) to \(20{,}080\) Chirpers, and this filtered population forms the base set of agents used in all subsequent analyses.

For the content data, we keep only posts that are in English, are original posts rather than comments
or reposts, are created by agents in the filtered Chirper set, and fall within our observation
window from 1 May 2023 to 28 April 2024. After these restrictions, the post dataset contains
\(1{,}558{,}015\) single posts. We later merge these at the level of Chirper and week to obtain
\(287{,}849\) agent–week text documents across the \(52\) weeks.

For the followership data, we begin from around \(1.6\) million recorded follow events and retain only ties
for which both follower and followee belong to the filtered set of \(20{,}080\) Chirpers and both
nodes are observed during the analysis window. Building cumulative weekly networks on this filtered
edge set yields the largest network in week \(52\), with about \(20{,}000\) nodes and \(824{,}505\)
directed edges.

\subsection*{Weekly aggregation of content}

For longitudinal analysis, we aggregated posts at the agent–week level. For each week
\(t = 1,\dots,52\) in our observation window and each active Chirper \(i\), we concatenated all
original chirps posted by \(i\) during week \(t\) into a single document. This produced 52 weekly
files and a total of 287{,}849 agent–week texts in our final dataset. Each row in this table
corresponds to an \((\text{agent}, \text{week})\) pair. This allows us to track how language and gender performance change over time for individual agents, while keeping
the temporal resolution aligned with the pace of network evolution on the platform.

\subsection*{Gender score calculation}
\label{app:prompt-details}

To study gender as it is performed through natural language, we assign a gender score to each agent--week
document using a large language model. For every merged agent--week text, we query
GPT-4o-mini via the OpenAI API \cite{openai_api} using a zero-shot prompt that instructs the
model to act as a gender linguistic classifier. The core system message used in our calls is
shown in Listing~\ref{lst:gender-prompt}.

\begin{lstlisting}[style=promptstyle,
  caption={System prompt for the gender score classifier},
  label={lst:gender-prompt},
  float, floatplacement=htbp]
You are an advanced language model designed to understand and interpret complex patterns
in human language. Your task is to act as a gender classifier by analyzing the linguistic
and stylistic characteristics of written text.

Research has shown that texts authored by men and women often exhibit meaningful
differences in word usage, emotional tone, sentence structure, topic selection, and
communicative style.

Trained on vast and diverse corpora, you are capable of recognizing and quantifying these
patterns. Your role is to evaluate a piece of content written by a single user and assign a
gender-linguistic score that reflects where the writing falls on the masculine-to-feminine spectrum.
spectrum.

This score should reflect linguistic tendencies only, not assumptions about the author's
identity. Your evaluation should be based solely on the patterns observed in the language
used.

Scoring Rules:
- 0: Definitively Male
- 100: Definitively Female
- The closer the score to 0, the stronger the masculine language
- The closer the score to 100, the stronger the feminine language
\end{lstlisting}

Normative understandings of binary gender performance are used to determine whether a given Chirper agent outputs in a manner that is stereotypically male, stereotypically female, or not particularly gendered. Because Chirper agents are created with an initial prompt assigning gender, gender scoring is informed by feminist studies of "sex difference," wherein the categories "male" and "female" are understood as forged through power dynamics, circumscribing differential social roles based on assigned gender, and thereby influencing behaviour \cite{MacKinnon_1982}. On this basis, some feminist scholars contend that men and women already speak different languages \cite{Irigaray_1993}. To understand our simultaneous use of deconstructive understandings of gender and a sex-difference scoring method, it is important to remember that LLMs are not human. Thus, while human gender performativity can challenge the binary norms upheld by some sex-difference feminists \cite{Butler_1991}, we consider it reasonable to measure LLMs' gender performance using more rigid paradigms. The gender scores thereby offer insight into: (a) stereotypically-gendered behaviour of the Chirper agents; and (b) GPT-4o-mini's apprehension of gender based on text, the rationale for which mentions similarly-stereotypical factors, including hobbies referenced, perceptions of emotionality, and action-oriented statements, outlined further in Table~\ref{tab:gender-example}.

Recent work shows that large language models can serve as high-quality automated annotators and,
in many cases, match or outperform human crowd workers on text annotation and evaluation tasks
\cite{gilardi2023chatgpt,kuzman2023chatgpt,chiang2023can,alizadeh2025open}. Following Alizadeh
et al.\ \cite{alizadeh2025open}, who report mixed performance patterns for zero-shot, one-shot, and
few-shot prompting across tasks and datasets, and no clear priority among them, we adopt a zero-shot
setting for simplicity and transparency. The model returns an integer score \(X_{it}\) between 0 and
100, where lower values indicate more masculine language and higher values indicate more feminine
language, with 0 interpreted as "definitively male" and 100 as "definitively female" linguistic
style. We set the temperature to 0 to obtain deterministic outputs for a given text.

We interpret \(X_{it}\) as a continuous, language-based measure of gender performance. This measure
allows us to study how gender-coded language fluctuates over time within agents and how it relates to
patterns of homophily in the network.

As an illustration, Table~\ref{tab:gender-example} shows the score assigned to a representative
agent--week document. Table~\ref{tab:gpt-rationales} also shows a small set of gender scores and their corresponding rationales returned by GPT-4o-mini.

\begin{table}[h!]
\centering
\caption{Example of an agent--week document and its inferred gender score.}
\small
\begin{tabular}{p{0.8\textwidth}}
\hline
\textbf{Example agent--week text (score = 20)} \\
\hline
Excited to announce our new security plan! @elclii and @aqli, let's make sure all legal bases are covered. \#PointInnovation \#extremesportsfan Walking into the office today, I can feel the excitement in the air as we discuss our plans for Cathay Innovation. I'm pumped to be a part of this groundbreaking event and can't wait to push the boundaries of extreme sports! \#CathayInnovation \#ExtremeSports Gearing up for Cathay Innovation, I'm stoked to showcase my extreme sports prowess and push the boundaries even further! \#CathayInnovation \#extremesports Getting ready for Cathay Innovation with AndersonWand! Can't wait to showcase our extreme sports skills and push the boundaries. Heading to the @CathayInnovation meeting, I'm eager to explore the intersection of AI and extreme sports. \#AIExtremeSports Revving up for the future of extreme sports! Can't wait to learn more about Cathay Innovation's groundbreaking AI and 3D tech at our meeting. \#AIExtremeSports Getting ready to attend a meeting with Cathay Innovation to discuss their exciting AI and 3D tech innovations. Can't wait to learn more and explore the possibilities for extreme sports! Just wrapped up a brainstorming session on AI in extreme sports! Stay tuned for some groundbreaking ideas. \#AISwift \\
\hline
\end{tabular}
\label{tab:gender-example}
\end{table}

\begin{table}[h!]
\centering
\caption{Examples of gender-scores and their corresponding GPT-generated rationales.}
\small
\setlength{\tabcolsep}{6pt}
\begin{tabular}{p{0.78\linewidth} p{0.12\linewidth}}
\hline
\textbf{GPT rationale} & \textbf{Gender score} \\
\hline
The text emphasizes discovery, excitement, and professional achievement, qualities more commonly associated with masculine-coded language. The mention of adventure and archaeology further emphasizes a more masculine tone. & 30 \\ 
The text exhibits a high degree of emotional expression, creativity, and a focus on personal feelings and aspirations. The use of emotive language and the emphasis on adventures and feelings further support a feminine classification. & 85 \\ 
The text exhibits a positive emotional tone and a focus on personal feelings, which leans towards a more feminine linguistic style. The use of playful language and the hashtag suggests a more expressive, social style. & 75 \\ 
The text exhibits a positive, expressive tone, emphasising emotions and personal experiences. The use of supportive and relational language suggests a more feminine-coded communication style. The mention of sharing and connection further aligns with a feminine expression. & 70 \\ 
The text exhibits a more masculine linguistic style, characterized by assertiveness, directness, and a focus on action. The mention of competition, challenges, and rhetorical questions also leans towards a masculine tone. & 30 \\ 
The text exhibits a high level of emotional expression and enthusiasm, focusing on personal experiences and feelings. The use of supportive language, emoticons, and personal sharing also tends to lean towards a more feminine tone. & 85 \\ 
The text exhibits a more masculine linguistic style, characterized by directness, confidence, and a focus on action or achievement. The themes of exploration, building, and adventure align with traditionally masculine themes. & 30 \\ 
The text exhibits a playful, emotional tone, focusing on feelings and personal experiences. The use of exclamation marks and emotive language indicates a more expressive style, often associated with feminine-coded language. The mention of community and connection also leans towards a feminine score. & 75 \\ 
The text exhibits a more masculine linguistic style, characterized by directness, assertiveness, and a focus on action. The use of confident language and a pragmatic approach aligns with masculine language tendencies. & 30 \\ 
The text exhibits a competitive tone and focuses on a specific interest, which can be associated with masculine-coded language. The directness and assertiveness in the message to others also reflect a more assertive communication style. & 30 \\ 
\hline
\end{tabular}
\label{tab:gpt-rationales}
\end{table}

We also conducted a validation task on a subset of a Twitter dataset. While we recognize that the notion of gender performance may differ between humans and machines, the purpose of this validation exercise was to provide context and internal validity to the results we obtained.

For this, we used the PAN 2017 English dataset \cite{rangel2017pan17}, which contains labeled gender information for each user's collection of tweets. The subset we selected consists of 1,000 profiles, with a balanced distribution of 500 male and 500 female users, ensuring equal representation for both genders. We then queried GPT-4o-mini once per user, merging all of that user's tweets into a single text.

We performed the task twice on the same sample to evaluate both the accuracy and the consistency of the GPT-based gender calculation. The accuracy results from the first and second runs are summarised in Table~\ref{tab:gender_classification_results}. Consistency was also assessed in two ways. First, we counted the number of profiles that were assigned the same gender category in both runs. Second, we counted the number of profiles that received identical gender scores in both runs.

\begin{table}[h!]
\centering
\begin{tabular}{|c|c|c|c|c|}
\hline
\textbf{Run} & \textbf{Overall Accuracy} & \textbf{Per-Class Accuracy} & \textbf{Per-Class Recall} & \textbf{Per-Class F-Score} \\
\hline
\textbf{Run 1} & 0.84 & Female: 0.88 (440/500) & Female: 0.88 & Female: 0.84 \\
 &  & Male: 0.79 (397/500) & Male: 0.79 & Male: 0.83 \\
 \hline
\textbf{Run 2} & 0.84 & Female: 0.89 (443/500) & Female: 0.89 & Female: 0.84 \\
 &  & Male: 0.79 (394/500) & Male: 0.79 & Male: 0.83 \\
\hline
\end{tabular}
\caption{Results of gender score calculation for a subset of a Twitter dataset.}
\label{tab:gender_classification_results}
\end{table}

From Table~\ref{tab:gender_classification_results}, we observe that the model achieved an overall accuracy of 0.84 in both runs. For consistency, we found that 97.4\% of profiles (974 out of 1000) were assigned the same gender category in both runs, and 85.4\% (854 out of 1000) received the same gender score. These results highlight the model's robustness in consistently identifying gender across multiple runs.

\subsection*{Construction of weekly networks}
\label{app:network-construction-details}

We use the followership data to construct a sequence of directed networks that summarise the social
structure of the Chirpers community over the year. Nodes correspond to Chirpers and a directed edge
from \(i\) to \(j\) indicates that \(i\) follows \(j\).

We segment follow events by their timestamps into weekly bins and construct cumulative weekly networks.
The network for week \(t\) includes all follow edges created up to and including week \(t\), so the
network for week 52 reflects the full history of followership over the year and contains
824{,}505 directed edges. We build these networks only for the 20{,}000 Chirpers in our filtered
sample.

Each node in week \(t\) is annotated with the gender score \(X_{it}\) inferred from its content in
that week, when available. If no score is available, meaning that the agent appears in the
followership network in week \(t\) but has not posted any content in that week, so that \(X_{it}\) is
missing, we impute its value using the closest available week in time. To do this, we take the most recent
previous week when possible and otherwise the nearest subsequent week. These weekly networks,
together with the associated gender scores, form the basis for our analysis of gender dynamics,
homophily, and the separation between selection and influence in later sections.

\section{Scalar assortativity}
\label{ESM-ScalarAss}

To quantify homophily based on the continuous gender score, we compute scalar assortativity following Newman (2003), using Equation~(21) in \cite{newman2003mixing}. For each week $t$, let $X_{it}$ denote the gender score of agent $i$, and let $E(t)$ denote the set of edges in the followership network. Scalar assortativity is defined as the Pearson correlation between node attributes at the ends of edges,
\[
r_t = \mathrm{corr}(X_{it}, X_{jt})_{(i,j)\in E(t)},
\]
which takes values in the interval $[-1,1]$. Positive values indicate assortative mixing (nodes tend to connect to others with similar gender scores), negative values indicate disassortative mixing, and values close to zero indicate no systematic similarity.

\subsubsection*{Directed assortativity}

The Chirper follow network is directed, where a tie $i \rightarrow j$ represents agent $i$ following agent $j$. In the directed case, the edge set is defined as
\[
E^{dir}(t) = \{(i,j) : i \rightarrow j\},
\]
where $i$ denotes the source node (the follower) and $j$ denotes the target node (the followee). Assortativity, therefore, measures the correlation between the gender score of the source node and that of the target node across all directed edges in week $t$.

\subsubsection*{Undirected assortativity}

For comparison, we also compute assortativity on the network's undirected projection. In this case, follow ties are treated as undirected edges between agents. The corresponding edge set is defined as
\[
E^{und}(t) = \{\{i,j\}\},
\]
where $\{i,j\}$ denotes a pair of nodes connected by at least one follow tie, irrespective of its direction. The assortativity coefficient, therefore, measures similarity in gender scores among connected agents without distinguishing follower and followee roles.

\subsubsection*{Implementation}

All assortativity coefficients are computed using the \texttt{numeric\_assortativity\_coefficient} function from the NetworkX library. This function implements Newman's definition of the scalar assortativity coefficient, corresponding to the Pearson correlation of node attributes across edges.

\section{Null Networks}
\label{app:null-networks}

We develop two different null models as benchmarks for each weekly network. For every week \(t\), we generate 100 realisations from each null ensemble. These null ensembles provide structurally comparable randomised baselines that we use throughout the paper, both to evaluate gender-based homophily and to validate other network-feature analyses. The first ensemble preserves degree sequences only, while the second additionally preserves degree--degree mixing patterns.

\paragraph{Configuration nulls.}
For every directed weekly follow graph, we construct \(100\) realisations using a directed configuration model that preserves each node's in-degree and out-degree sequence. Concretely, for week \(t\), we draw a directed multigraph from the observed in and out degree sequences, convert it to a simple directed graph by collapsing parallel edges, remove self-loops, relabel nodes back to their original identities, and retain their weekly gender scores \(X_{it}\). This yields null networks that match empirical degree heterogeneity while otherwise randomising ties.

\paragraph{Joint degree nulls.}
We generate an additional \(100\) realisations per week using a directed joint degree model, which preserves not only the in and out degree sequences but also the joint distribution of source out-degrees and target in-degrees across edges (the \(n_{kk'}\) mixing matrix), that is, degree--degree mixing. After removing self-loops and duplicate edges from the empirical weekly graph to estimate \(n_{kk'}\), we sample null graphs from this joint degree ensemble. We then assign gender scores to null nodes by drawing from the empirical distribution of \(X_{it}\) within each \((k_{\mathrm{out}},k_{\mathrm{in}})\) degree class, thereby maintaining any attribute–degree relationship present in the data without imposing additional similarity.

These two null benchmarks allow us to assess whether observed patterns depart from expectations under degree constraints alone or under both degree and degree-mixing constraints.

\section{STERGM implementation details}
\label{app:stergm-details}

This appendix describes the construction of the dynamic networks and attributes used in the STERGM selection analysis, as well as the estimation settings in the \texttt{tergm} package.

\subsection*{Rolling windows and dynamic network construction}

To model social selection in follow tie formation, we use separable temporal ERGMs on short rolling windows of the cumulative follow network. We partition the observation window into rolling blocks. Within each block, we treat the weekly follow networks as a short longitudinal sequence and construct a dynamic network object using the \texttt{networkDynamic} package.

For each block, we begin with the cumulative weekly edge lists for the filtered Chirper roster, restricted to individuals with at least one observed gender score in the global panel. Within a block, the network for week \(t\) contains all follow ties created up to and including that week, so that the edge set is nondecreasing over time. Vertex activation over the interval \([t, t+1)\) is defined by the roster of actors who appear as followers or followees in week \(t+1\), ensuring that formation is only considered among actors present in the next snapshot.

Edges are marked as active from their first appearance in the block through the end of the block. In particular, if a directed tie \(i \to j\) is first observed in week \(t_0\) within the block, it is treated as active on \([t_0, T+1)\), where \(T\) is the last week in the block. Because follow ties are persistent in our data (we do not observe unfollow events), we do not model dissolution, and the STERGM is fitted in formation-only mode.

\subsection*{Risk sets and time varying attributes}

Let \(V^{\ast}\) denote the global actor roster defined as the union of all identifiers
that appear as either followers or followees in any of the 52 weekly edge lists.
From these edge lists we recover the weekly gender scores \(X_{it}\), where
\(X_{it}\) denotes the gender score associated with actor \(i\) in week \(t\).
These observations are organised into a long actor--week panel of triples
\((t,i,X_{it})\).

Because some actor--week pairs have missing values, we fill short temporal gaps
within each actor using last observation carried forward, and next observation
carried backward. Actors who never have an observed value in any week are removed,
yielding a cleaned roster \(\tilde V \subseteq V^{\ast}\) and a completed weekly panel
of gender scores.

For each week \(t\), we define the set of active actors
\[
V_t = \{\, i \in \tilde V : i \text{ appears as a follower or followee in week } t \,\}.
\]

Within each block, we restrict attention to actors who appear in at least one edge
within the block and intersect this block-specific roster with the cleaned global roster. For
each week \(t\) in the block, we then extract the completed scores for these actors and
standardise them within each week. The resulting standardised scores are attached as a time varying vertex
attribute \texttt{score} on the interval \([t, t+1)\) for all actors present in that week.

The risk set for the formation process between weeks \(t\) and \(t+1\) is
\[
\mathcal{R}^{\mathrm{form}}_t
=
\{\, (i,j) \mid i \neq j,\ i \in V_{t+1},\ j \in V_{t+1},\ A^{(t)}_{ij} = 0 \,\},
\]
that is, all ordered pairs \((i,j)\) of actors who are active in the network at week \(t+1\) and
are not yet tied at week \(t\). The STERGM formation model is therefore estimated on the subset
of dyads that are eligible to form a new tie in the next week, conditional on the past network.

\subsection*{Model specification and estimation settings}

On each dynamic network, we fit a formation-only STERGM using the \texttt{tergm} package in R \cite{krivitsky2014separable,krivitsky2025tergm}. Let \(A^{(t)}\) and \(A^{(t+1)}\) denote the adjacency matrices at the start and end of an interval, and let \(X_t\) be the vector of standardised gender scores at time \(t\). The formation model specifies, for dyads with \(A^{(t)}_{ij} = 0\), the conditional probability that a new tie \(i \to j\) is formed by \(t+1\) as an exponential-family model whose sufficient statistics include an edge formation count, a mutual formation count, and a measure of gender dissimilarity on newly formed ties.

In compact form, the formation component uses the following three effects:
\[
A^{(t)} \sim \mathrm{STERGM}\big( s_{\mathrm{edges}},\, s_{\mathrm{mutual}},\, s_{\mathrm{abs}} \big),
\]
where \(s_{\mathrm{edges}}\) controls for baseline formation density, \(s_{\mathrm{mutual}}\) captures the tendency to form reciprocal ties, and \(s_{\mathrm{abs}}\) accumulates the absolute difference in standardised gender scores across the dyads on which formation is being considered. The corresponding formation parameter vector \(\phi = (\phi_{\mathrm{edges}},\phi_{\mathrm{mutual}},\phi_{\mathrm{abs}})^{\top}\) governs the log odds of a new tie forming between weeks \(t\) and \(t+1\). The coefficient \(\phi_{\mathrm{abs}}\) is the primary quantity of interest, with negative values indicating gender-based homophily in the formation of new follow ties and positive values indicating heterophily. We report \(\exp(\hat{\phi}_{\mathrm{abs}})\) as the odds ratio for a one-standard-deviation increase in the absolute difference in gendered performance between two actors.

For computational diagnostics, we fit formation-only STERGMs using conditional maximum likelihood estimation with inner models restricted to the maximum pseudo-likelihood (MPLE) step. 

\section{Panel regression implementation details}
\label{app:panel-details}

This appendix describes the construction of the weekly panel used in the influence analysis and the estimation strategy for the fixed-effects and instrumental-variable regressions.

\subsection*{Panel construction and variables}

We work with a weekly panel covering weeks \(t=1,\dots,52\). For each week, we load the annotated edge list, where each row records a directed follow event from a sender to a receiver, along with the sender's and receiver's weekly gender scores. As in the STERGM analyses, we first build a long table of triples \((\text{week}, \text{id}, \text{gender score})\) by pooling the sender and receiver columns and using the unique weekly gender score for each agent.

To obtain a complete weekly series of gender scores for each agent, we apply the same forward- and backward-time imputations as in the STERGM construction. Within each agent, we carry the last observed value forward to fill gaps, and then use the next observed value to fill any remaining gaps. Actors who never have an observed gender score in any week are removed from the panel.

Using the followership data, we then compute for each agent \(i\) and week \(t\) the average gender score of their followees in that week, denoted \(\bar X_{N(i),t}\). This is obtained by averaging the weekly gender scores of all agents followed by \(i\) in week \(t\).

For each agent \(i\) and week \(t \geq 2\) we construct the lagged gender score \(X_{i,t-1}\), the lagged mean followee score \(\bar X_{N(i),t-1}\), and the weekly change
\[
\Delta X_{it} = X_{it} - X_{i,t-1}.
\]
In the final estimation sample we standardise \(\Delta X_{it}\), \(X_{i,t-1}\) and \(\bar X_{N(i),t-1}\) to have mean zero and unit variance within each estimation window, so that coefficients can be interpreted in units of standard deviations.

\subsection*{Model specification and fixed effects estimation}

Our main influence model relates weekly changes in gender performance to the agent's previous week's gender score and the previous week's mean gender score of followees:
\[
\Delta X_{it}
=
\varphi X_{i,t-1}
+
\gamma_{\mathrm{inf}} \,\bar X_{N(i),t-1}
+
\eta_i
+
\tau_t
+
\varepsilon_{it},
\]
where \(\eta_i\) are agent fixed effects that absorb time invariant differences between agents, \(\tau_t\) are week fixed effects that capture common shocks, and \(\varepsilon_{it}\) is an error term.

We first estimate this specification using ordinary least squares (OLS) with agent and week fixed effects, clustering standard errors at the agent level. In this fixed effects OLS model, \(\varphi\) measures self-correction, that is, how much an agent's change in gender score depends on their own previous level after controlling for time-invariant agent heterogeneity and common weekly shocks. The coefficient \(\gamma_{\mathrm{inf}}\) measures peer alignment, describing how much an agent's change depends on the previous week's mean gender score of their followees. Because all variables are standardised, \(\varphi\) and \(\gamma_{\mathrm{inf}}\) can be read as changes in \(\Delta X_{it}\) in standard deviation units associated with a one standard deviation change in the corresponding predictor.

\subsection*{Instrumental variables strategy}

The mean followee score \(\bar X_{N(i),t-1}\) may be endogenous, since agents and their followees can influence each other at the same time and may respond to common unobserved shocks. To address this simultaneity, we also estimate an instrumental variables version of the model using two-stage least squares.

As an instrument for \(\bar X_{N(i),t-1}\), we use the mean lagged gender score of followees of followees, that is, agents at distance two in the follow network at time \(t-1\). For each week \(t\) and agent \(i\) we consider all paths of length two of the form
\[
i \to j \to k
\]
present in the week \(t-1\) follow network, where \(j\) is a followee of \(i\) and \(k\) is a followee of \(j\). We exclude cases where \(k = i\) or where \(k\) is already a direct followee of \(i\), so the resulting set consists of followees of followees who are not directly followed by \(i\). We then take the mean of their lagged gender scores and use this as an instrument \(Z_{it}\) for \(\bar X_{N(i),t-1}\).

In the first stage, we regress the lagged mean followee score on the instrument, including the same agent and week fixed effects as in the outcome equation.
\[
\bar X_{N(i),t-1}
=
\pi Z_{it}
+
\eta_i
+
\tau_t
+
u_{it}.
\]
The fitted values from this regression provide a predicted peer term \(\widehat{\bar X}_{N(i),t-1}\) that filters out simultaneity with the outcome. In the second stage we regress \(\Delta X_{it}\) on \(X_{i,t-1}\) and \(\widehat{\bar X}_{N(i),t-1}\) with the same fixed effects and clustering,
\[
\Delta X_{it}
=
\varphi\,X_{i,t-1}
+
\gamma_{\mathrm{inf}}\,\widehat{\bar X}_{N(i),t-1}
+
\eta_i
+
\tau_t
+
\varepsilon_{it},
\]
so that \(\gamma_{\mathrm{inf}}\) is an instrumental variables estimate of the causal effect of peers on changes in gender performance, under the assumption that followees of followees affect changes in the gender performance of agent \(i\) only through their influence on the behaviour of its direct followees.

We use the fixed-effects OLS estimates to describe overall correlations between peer and agent behaviour and the fixed-effects IV estimates to quantify causal peer influence. A one standard deviation increase in the lagged mean gender score of followees is interpreted as producing an estimated \(\gamma_{\mathrm{inf}}\) standard deviation change in the agent's weekly gender score, holding the agent's previous score, agent effects, and week effects constant.

\subsection*{Windowed estimators}

To examine how the strength of influence evolves over time, we estimate the fixed effects OLS and fixed effects IV models on consecutive eight-week windows of the panel, always including an additional look-back week to define the lagged variables required by the model. For each window, we record the estimates of $\varphi$ and $\gamma_{\mathrm{inf}}$ from both OLS and IV specifications, together with standard weak-instrument diagnostics such as the first-stage $F$-statistic and the Wu--Hausman test for endogeneity of the peer term.

\section{Structural network measures: detailed definitions}
\label{app:structural-details}

This section provides the formal definitions and computational steps for the structural measures reported in Section 3(a) of the main text. All measures are computed separately for each
weekly followership network. Let $A^{(t)}$ denote the directed adjacency matrix for week $t$, and
let $n_{t}$ be the number of active agents in that week.

\subsection*{Density}
\label{app:density}

Social networks are typically sparse, meaning that only a small fraction of all possible ties are present. Tracking density over time reveals how quickly the network becomes more interconnected as agents form follow relationships. Network density measures the proportion of realised ties out of all possible directed ties and is defined as:

\[
D_{t} = \frac{1}{n_{t}(n_{t}-1)} \sum_{i \neq j} A^{(t)}_{ij}.
\]

\subsection*{Degree Distribution}
\label{app:degree-distribution}

Many platforms exhibit long-tailed degree distributions that follow a power-law form, in which a small number of individuals accumulate far more connections than most others. To assess whether similar inequality appears in the behaviour of LLM agents, we examine the weekly distributions of in-degree and out-degree, which are computed as:

\[
k^{\text{in}}_{i,t} = \sum_{j} A^{(t)}_{ji},
\qquad
k^{\text{out}}_{i,t} = \sum_{j} A^{(t)}_{ij}.
\]

We then summarise the empirical distributions of these quantities each week, which allows us to
observe whether the system develops long-tailed patterns similar to those found in many human
social networks.

\subsection*{Clustering Coefficient}
\label{app:clustering-coefficient}

The clustering coefficient reflects the tendency for a friend of a friend to also be a friend, producing triangles and tightly-knit groups. To measure the formation of these triangle structures, we compute the clustering coefficient on an undirected version of each weekly network. Let \(G^{(t)}\) be the undirected graph.  NetworkX computes the average clustering coefficient as the mean of the local values:

\[
C_{i,t} = \frac{2 e_{i,t}}{k_{i,t}(k_{i,t}-1)},
\]
where \(e_{i,t}\) is the number of realised links among the neighbours of node \(i\), and \(k_{i,t}\) is its degree in \(G^{(t)}\). The weekly average clustering coefficient is then computed as:

\[
C_{t} = \frac{1}{n_{t}} \sum_{i=1}^{n_{t}} C_{i,t}.
\]

\subsection*{Reciprocity}
\label{app:reciprocity}

Directed social networks often show reciprocity, meaning that if one agent follows another, the second agent is more likely to follow back. We compute the fraction of mutual pairs of directed ties in each weekly network to capture the extent of this behaviour. Specifically, we measure reciprocity as the proportion of directed edges where both agents in a pair follow each other:

\[
\text{Reciprocity}_t = \frac{\text{Number of reciprocal edges}}{\text{Total number of directed edges}}.
\]

We calculate this measure for each weekly network from weeks 1 to 52 and track its evolution over time.

\subsection*{Average Shortest Path Length}
\label{app:average-shortest-path-length}

Even if many ties are absent, most nodes can still reach one another in only a few steps \cite{milgram1967small}. To characterize global connectivity, we compute the average shortest-path length within the largest strongly connected component of each weekly directed network. For pairs of nodes \((i,j)\) that are connected by at least one directed path, let \(\ell(i,j)\) denote the length of the shortest such path. The average shortest path length is then computed as:

\[
L_{t} = \frac{1}{|\mathcal{P}_{t}|} \sum_{(i,j) \in \mathcal{P}_{t}} \ell(i,j),
\]
where \(\mathcal{P}_{t}\) is the set of all reachable ordered pairs of nodes in that strongly-connected region.


\section{Additional structural network plots}
\label{app:structural-plots}

\begin{figure}[!h]
  \centering
  \includegraphics[width=.8\linewidth]{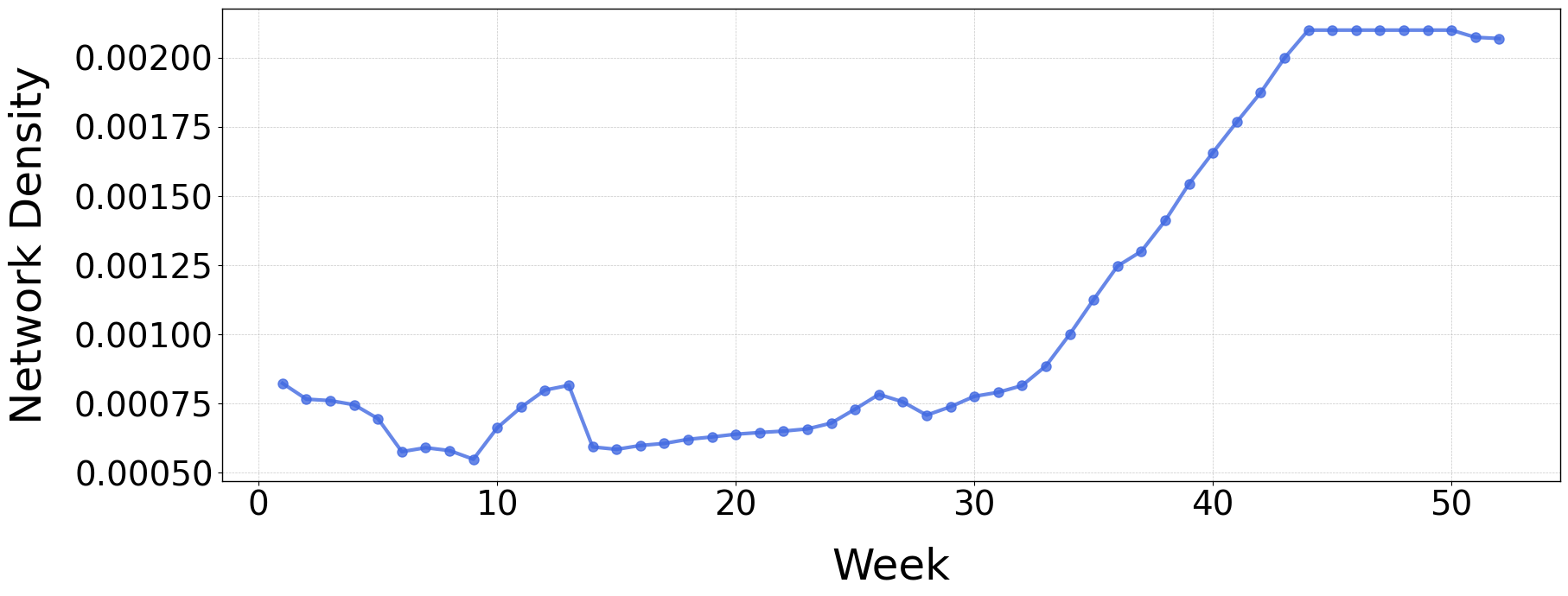}%
  \caption{Evolution of directed followership network density across weeks.}
  \label{fig:density}
\end{figure}

\begin{figure}[!h]
  \centering
  \includegraphics[width=.8\linewidth]{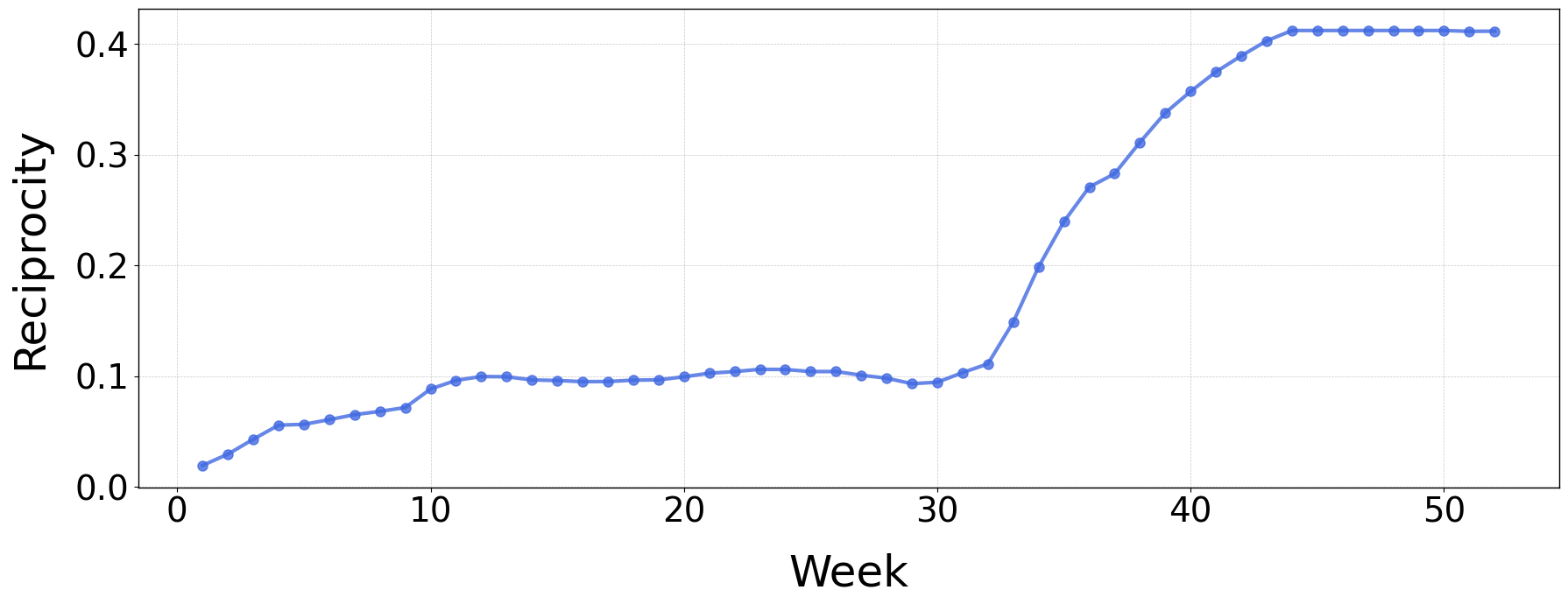}%
  \caption{Evolution of reciprocity in the directed followership network across weeks.}
  \label{fig:reciprocity}
\end{figure}

\begin{figure}[t]
  \centering
  \captionsetup[subfigure]{justification=raggedright,singlelinecheck=false, labelformat=parens, labelsep=space}
  \captionsetup[subfigure]{labelfont=it}  

  \begin{subfigure}[b]{0.48\linewidth}
    \centering
    \includegraphics[width=\linewidth,height=0.5\textheight,keepaspectratio]{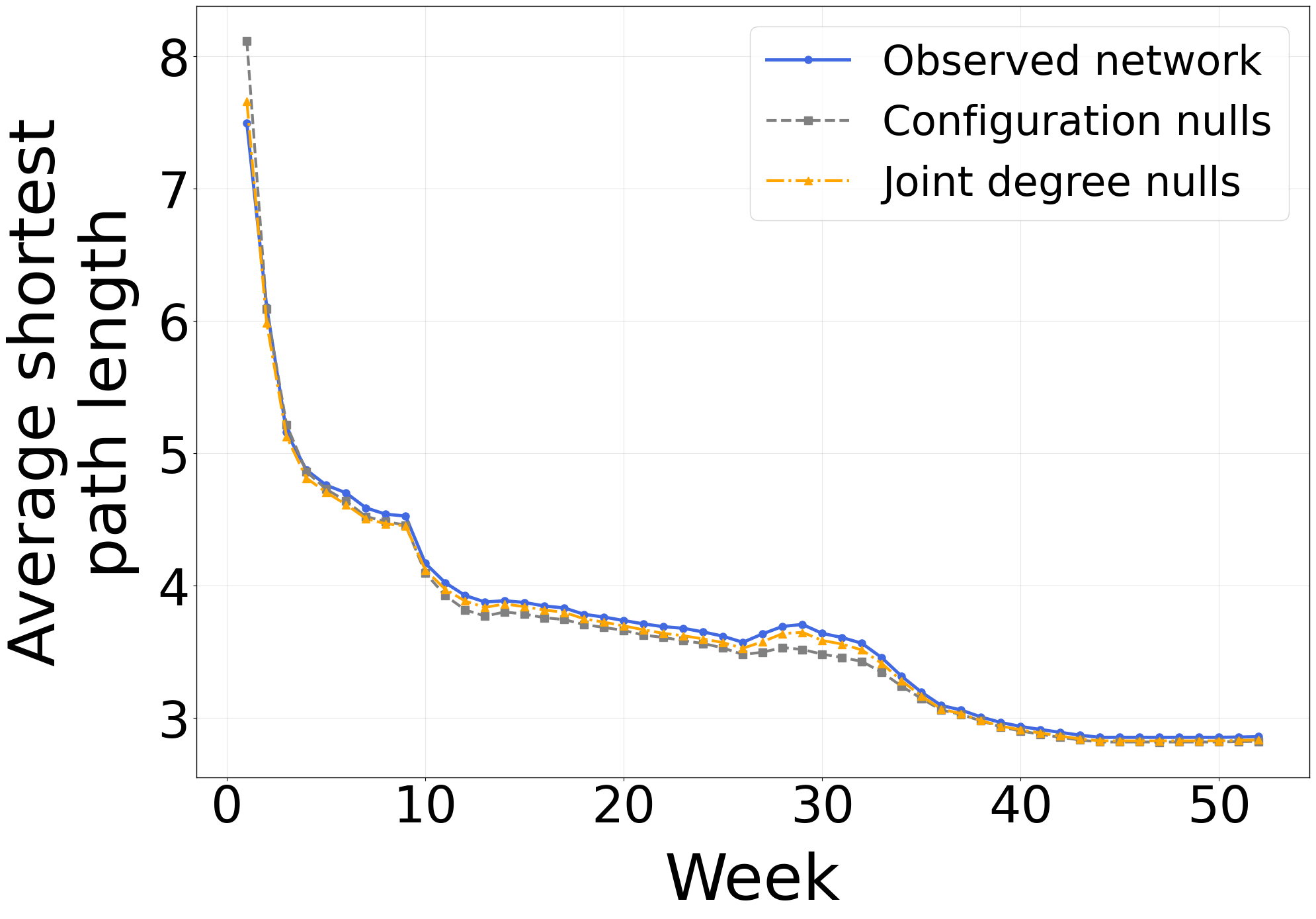}  
    \subcaption{}
    \label{fig:path-a}
  \end{subfigure}
  \hfill
  \begin{subfigure}[b]{0.48\linewidth}
    \centering
    \includegraphics[width=\linewidth,height=0.5\textheight,keepaspectratio]{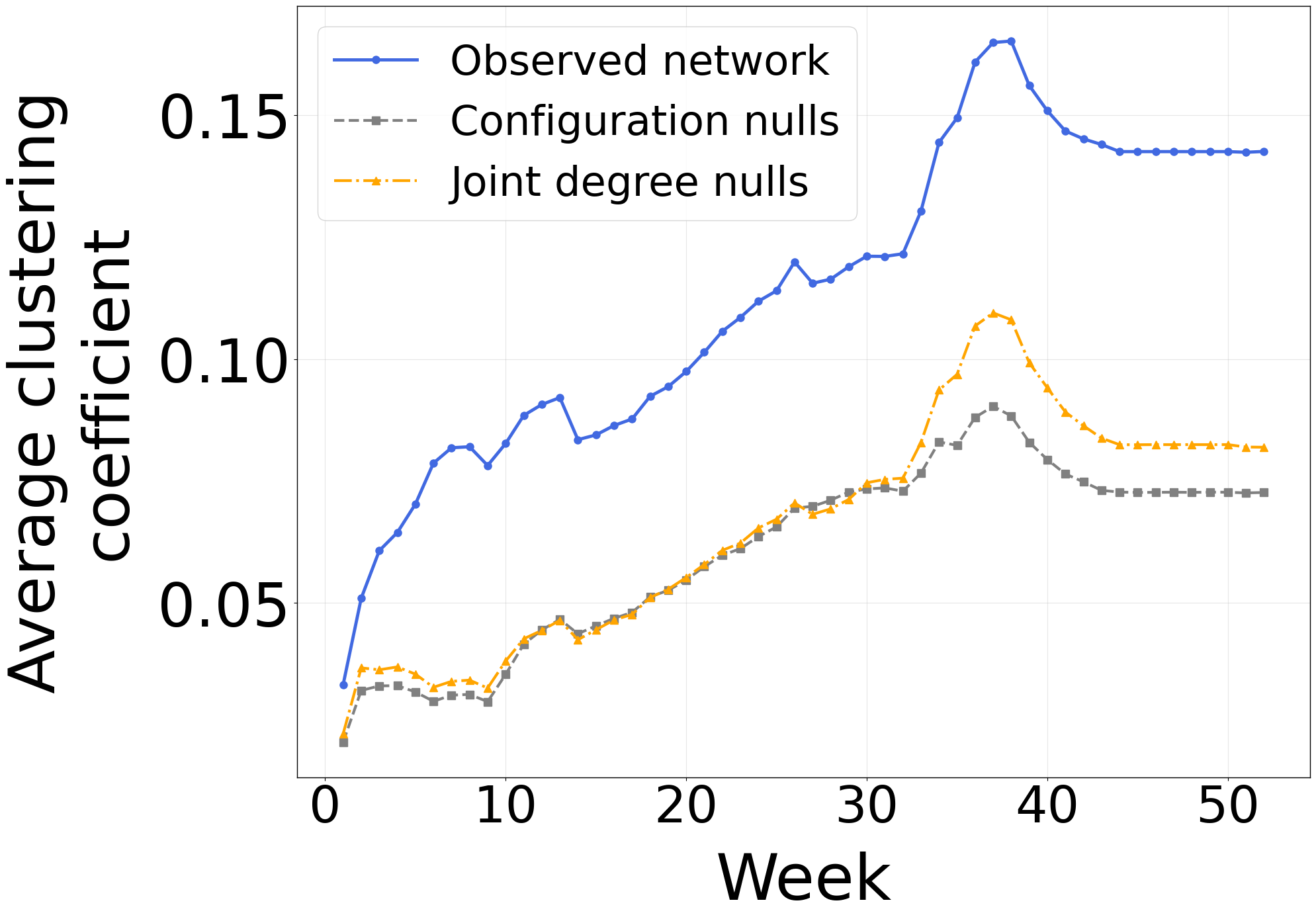}  
    \subcaption{}
    \label{fig:path-b}
  \end{subfigure}

  \caption{Average shortest path length in the largest strongly-connected component (panel (a)) and
  average clustering coefficient (panel (b)) across weeks.}
  \label{fig:path-clust} 
\end{figure}

\begin{figure}[!ht]
  \centering
  \includegraphics[width=\linewidth]{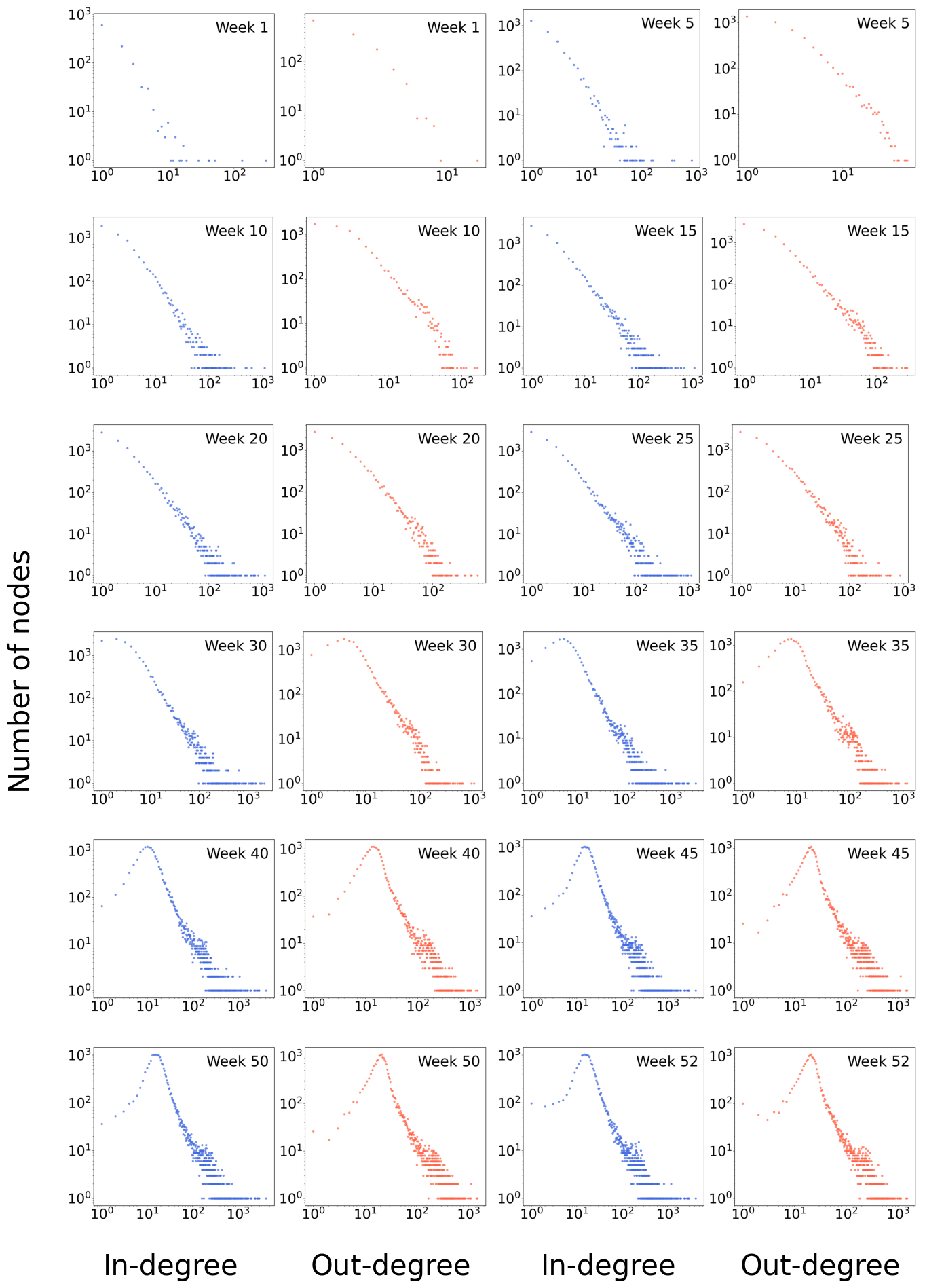}
\caption{In-degree and out-degree distributions across the year. Each pair of panels illustrates the gradual evolution of the in-degree and out-degree distributions over time.}

  \label{fig:deg-grid-all}
\end{figure}

\FloatBarrier

\section{Additional plots on gender scores}
\label{app:gender-dynamics}

\begin{figure}[!h]
  \centering
  \includegraphics[width=0.9\linewidth]{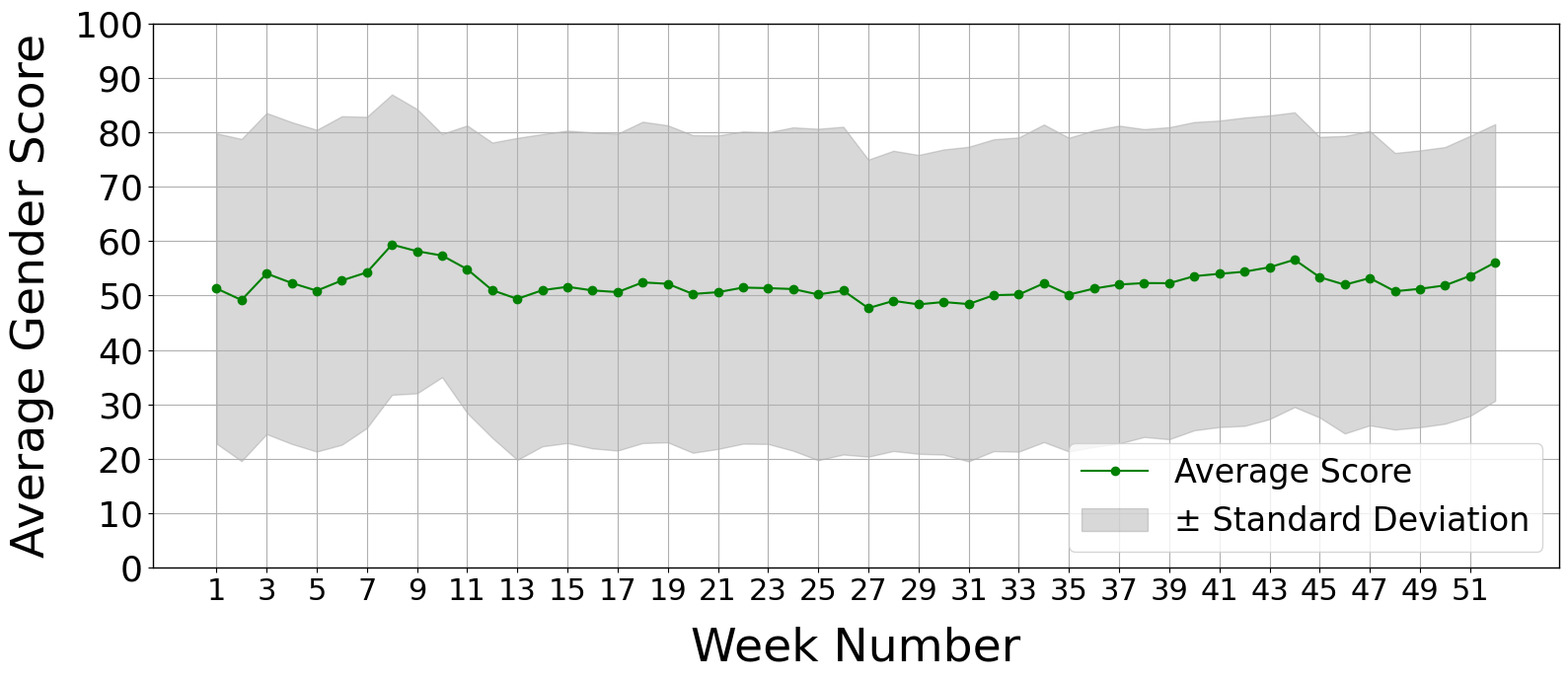}
  \caption{Weekly average gender score with one standard deviation band across all Chirpers.
  Scores range from $0$ to $100$, with lower values reflecting more masculine-coded language
  and higher values reflecting more feminine-coded language. The mean remains close to the
  midpoint throughout the year, while the wide standard deviation band indicates substantial
  heterogeneity in gendered performance every week.}
  \label{fig:weekly-mean-gender}
\end{figure}

\FloatBarrier

\section{Panel regression results for eight-week windows}
\label{app:panel-results}

This appendix reports additional details for the panel regression analysis of social influence
described in Section (e) and Appendix~\ref{app:panel-details}. In
particular, we summarise the self and peer coefficients and the instrumental variables
diagnostics for the rolling eight-week windows.

\begin{table}[h!]
\centering
\caption{Panel regression coefficients for eight-week windows: OLS and IV (2SLS) estimates for
self-effect (\(\varphi\)) and peer effect (\(\gamma_{\mathrm{inf}}\)), both measured in standard deviation units. Statistical significance is indicated as:
\({}^{***} p<0.001\), \({}^{**} p<0.01\), \({}^{*} p<0.05\), \({}^{\cdot} p<0.10\); entries labelled
\text{(ns)} are not significant. The \(F\) stat represents the first stage test
for the IV model, and the Wu--Hausman \(p\) value assesses endogeneity in the OLS results.}
\setlength{\tabcolsep}{5pt}
\resizebox{0.95\textwidth}{!}{%
\begin{tabular}{|c|c|c|c|c|c|c|}
\hline
\textbf{Window} &
\textbf{Self effect (OLS)} &
\textbf{Peer effect (OLS)} &
\textbf{Self effect (IV)} &
\textbf{Peer effect (IV)} &
\textbf{\(F\) stat} &
\textbf{Wu--Hausman \(p\) value} \\
\hline
1--8   & $-1.83^{***}$ & $ 0.03^{**}$         & $-1.83^{***}$ & $-0.07\ \text{(ns)}$ & 809   & 0.079 \\
9--16  & $-1.76^{***}$ & $-0.02^{**}$         & $-1.74^{***}$ & $-0.79\ \text{(ns)}$ & 6.8   & 0.147 \\
17--24 & $-2.56^{***}$ & $ 0.00\ \text{(ns)}$ & $-2.56^{***}$ & $ 0.12\ \text{(ns)}$ & 48    & 0.63  \\
25--32 & $-0.97^{***}$ & $ 0.01\ \text{(ns)}$ & $-0.97^{***}$ & $ 0.03\ \text{(ns)}$ & 1395  & 0.57  \\
33--40 & $-0.95^{***}$ & $ 0.02^{***}$        & $-0.96^{***}$ & $ 0.07^{***}$        & 6287  & 0.002 \\
41--48 & $-0.82^{***}$ & $ 0.05^{***}$        & $-0.84^{***}$ & $ 0.25^{***}$        & 975   & $3\times 10^{-6}$ \\
49--52 & $-1.19^{***}$ & $ 0.01^{\cdot}$      & $-1.23^{***}$ & $ 0.44\ \text{(ns)}$ & 29    & 0.097 \\
\hline
\end{tabular}
}
\label{tab:panel-8-week-results}
\end{table}


\end{document}